# MCMC with Adaptive Principal-Component Transformation: Rotation-Invariant Universal Samplers for Bayesian Structural System Identification

Xianghao Meng[1,2], Yong Huang[1,2,*], James L. Beck[3], Kui Jiang[4], Hui Li[1,2]

[1] Key Lab of Smart Prevention and Mitigation of Civil Engineering Disasters of the Ministry of Industry and Information Technology, Harbin Institute of Technology, Harbin, China

[2] Key Lab of Structures Dynamic Behavior and Control of the Ministry of Education, Harbin Institute of Technology, Harbin, China

[3] Division of Engineering and Applied Science, California Institute of Technology, CA, USA

[4] Faculty of Computing, Harbin Institute of Technology, Harbin, China

**Abstract**

Over decades, Markov chain Monte Carlo (MCMC) methods have been widely studied, with a typical application being the quantification of posterior uncertainties in Bayesian system identification of structural dynamic models. To address the issue of excessively low sampling efficiency in generic MCMC methods when applied to specific problems, researchers developed several MCMC algorithms that integrate trainable neural networks to replace and enhance their critical components. Later, meta-learning MCMC methods emerged to reduce training time. However, they require considerable similarity between test and training tasks, while their sampling efficiency is constrained by trade-off-simplified network designs. This paper proposes the Adaptive Principal-Component (PC) Meta-learning Stochastic Gradient Hamiltonian

Monte Carlo (APM-SGHMC) algorithm. It adaptively rotates coordinate axes in the parameter space to align with the PC directions of the current posterior samples, ensuring rotation-invariance of sampling performance with respect to the posterior distribution. By incorporating translation-invariance, scale-invariance, and rotation-invariance in a unified framework, APM-SGHMC enables universal samplers to acquire generalizable knowledge across diverse Bayesian system identification tasks using minimalistic tasks while eliminating the constraints imposed by network design trade-offs on sampling efficiency. Practical feasibility issues are also addressed. Two Bayesian system identification case studies demonstrate its effectiveness and universality: our method overcomes the case-by-case limitations of traditional data-driven approaches, achieving zero-shot generalization across structurally distinct models without retraining and maintaining consistent superior performance across all scenarios.

## 1. INTRODUCTION

Markov chain Monte Carlo (MCMC) methods, which are powerful computational tools, are widely studied for numerically sampling analytically complex posterior distributions [1-8] in Bayesian inference [9,10], typically in Bayesian system identification of structural models [11-18]. As a rational and robust framework, Bayesian inference utilizes Bayes' theorem to quantify the posterior uncertainty of parameter vectors from available data. Yet, multidimensional integrals in posterior calculation often render analytical solutions intractable. MCMC methods address this by generating samples as Markov chain states with a stationary distribution matching the posterior. Under the ergodicity assumption, posterior samples can be obtained through Markov chain simulation.

Classical MCMC algorithms, based on the Metropolis–Hastings (MH) algorithm, include transitional Markov chain Monte Carlo (TMCMC) [5-8], which handles large prior-posterior differences, and Hamiltonian Monte Carlo (HMC) [1,19-20], which accelerates state-space exploration by introducing auxiliary "momentum" variables and replacing the random-walk proposal with Hamiltonian dynamics. However, these generic methods are not customized for specific sampling problems. Further extension and optimization [21-23] help explore their potential for specific problems, but greater effectiveness demands more flexibility, leading to more tunable parameters and increased difficulty for practitioners.

In recent years, neural networks (NNs) have simplified the extension and optimization of generic MCMC methods. Many NN-enhanced MCMC algorithms [24-26] employ NNs to replace and improve their specific components for targeted problems, aiming for rapid convergence and efficient exploration during the sampling processes. However, these methods often suffer from reduced generalization ability after training, requiring retraining for new tasks, which is time-consuming and undermines competitiveness.

Meta-learning techniques [27-30] capture common knowledge from similar tasks through various ingeniously designed learning algorithms, ensuring generalization ability and enabling learners to handle a range of tasks with minimal training. Based on this concept, trained MCMC samplers can be directly applied to similar sampling tasks, thereby reducing training time and boosting the competitiveness of the NN-based algorithms.

A NN-enhanced SGHMC (NN-SGHMC) algorithm [26] replaces key extensible parts of the stochastic gradient HMC (SGHMC) algorithm with carefully designed component-wise NNs for training, ensuring meta-learning across tasks with varying dimensions, crucial for the generalization ability of a trained sampler. Its meta-learning potential has been demonstrated in some Bayesian deep learning tasks. Based on this work, we developed the adaptive meta-learning SGHMC (AM-SGHMC) algorithm [31] to ensure scale- and translation-invariance of the sampling performance with respect to the posterior probability density function (PDF), further enhancing generalization. The "adaptive" aspect stems from its adaptive moment estimation technique, extending that in Adam [34], to dynamically optimize the sampling strategy by continuously acquiring information about the target posterior PDF, consistent with other adaptive sampling approaches [3,18]. AM-SGHMC also modifies the loss function and its back-propagation path for more stable NN training in a Markov chain environment, enabling its use in structural dynamic model updating. However, its generalization capability remains limited. For instance, since their NN inputs consist solely of local component-wise information, and sampling strategies (NN outputs) are proposed separately for each component, these samplers struggle to fully explore high-probability regions when components are strongly correlated.

This paper proposes an adaptive principal-component (PC) meta-learning SGHMC (APM-SGHMC) algorithm. Within APM-SGHMC, building upon embedded scale- and translation-invariance properties, a novel adaptive PC transformation technique based on adaptive PC direction (PCD) estimation is introduced to ensure the rotation-invariance of the sampling performance with respect to the posterior PDF, thereby circumventing the limitations imposed by component-wise NN architectures on both sampler generalization and sampling efficiency, ultimately achieving universal generalization capabilities. Additionally, the adaptive estimation is further generalized to address the flexibility and accuracy requirements for estimated values across different sampling stages. Furthermore, a potential energy-based approach for convergence diagnostic and divergence mitigation is proposed to significantly boost robustness.

The paper is structured as follows: Section 2 outlines the challenges and framework of APM-SGHMC; Section 3 further details its main contributions; Section 4 presents the APM-SGHMC algorithm in pseudocode; Section 5 demonstrates its effectiveness and universal generalization capability via two case studies involving Bayesian inference of a building structural model and a bridge structural model; and Section 6 concludes with a summary.

## 2. APM-SGHMC: ADAPTIVE PRINCIPAL-COMPONENT META-LEARNING SGHMC FOR BAYESIAN STRUCTURAL SYSTEM IDENTIFICATION

In Bayesian system identification of structural models, the posterior PDF of the model parameter vector $\boldsymbol{w} \in \mathbb{R}^D$ is derived from Bayes' theorem:

$$p(\boldsymbol{w}|\boldsymbol{\mathcal{D}}) = p(\boldsymbol{\mathcal{D}}|\boldsymbol{w})p(\boldsymbol{w})/p(\boldsymbol{\mathcal{D}}) \tag{1}$$

where $p(\boldsymbol{w})$ is the prior PDF, $p(\boldsymbol{\mathcal{D}}|\boldsymbol{w})$ is the likelihood function for observed data $\boldsymbol{\mathcal{D}}$ and the evidence function $p(\boldsymbol{\mathcal{D}}) = \int p(\boldsymbol{\mathcal{D}}|\boldsymbol{w})p(\boldsymbol{w})\,d\boldsymbol{w}$ act as a normalizing constant ensuring the posterior PDF integrates to one.

For time-series data, the likelihood function is typically expressed as [1,31]:

$$p(\boldsymbol{\mathcal{D}}|\boldsymbol{w}) = \frac{1}{(2\pi\sigma^2)^{\frac{N_o N_T}{2}}} \times exp\left(-\frac{1}{2\sigma^2}\sum_{n=1}^{N_o}\sum_{j=1}^{N_T}[\hat{y}_n(t_j) - y_n(t_j; \boldsymbol{w})]^2\right) \tag{2}$$

Where $\hat{y}_n(t_j)$ and $y_n(t_j; \boldsymbol{w})$ denote measured and model-predicted outputs for the $n$th observed degree of freedom at time $t_j$, respectively. Errors are modeled as independent and identically distributed Gaussian variables with zero mean and some unknown variance $\sigma^2$, following the Principle of Maximum Entropy [32-33].

However, the analytical calculation of multidimensional integrals like the evidence function $p(\mathcal{D}) = \int p(\mathcal{D}|\boldsymbol{w})p(\boldsymbol{w})\,d\boldsymbol{w}$ becomes intractable for complex models. To address this, AM-SGHMC, the theoretical groundwork of APM-SGHMC, can be employed to approximate the posterior distribution through sampling.

## *2.1 AM-SGHMC: algorithm foundation and challenges*

AM-SGHMC is an MCMC method that generates Markov chains through stochastic particle dynamics simulations. In Bayesian inference, such MCMC methods treat the model parameter vector as particle positions $\boldsymbol{\theta} = \boldsymbol{w}$ in a state space, where the potential energy landscape $U(\boldsymbol{\theta}) = -\log(p(\mathcal{D}|\boldsymbol{\theta})) - \log(p(\boldsymbol{\theta})) + c^*$ is constructed from the unnormalized posterior PDF $p(\boldsymbol{\theta}|\mathcal{D}) \propto p(\mathcal{D}|\boldsymbol{\theta})p(\boldsymbol{\theta})$. This ensures the particles' stationary distribution $\pi(\boldsymbol{\theta}) \propto \exp(-U(\boldsymbol{\theta}))$ exactly matches the target posterior $\pi(\boldsymbol{\theta}) = p(\boldsymbol{\theta}|\mathcal{D})$, with $p(\boldsymbol{\theta})$ as the prior PDF and $p(\mathcal{D}|\boldsymbol{\theta})$ as the likelihood function for observed data $\mathcal{D}$.

In AM-SGHMC, the augmented state space is defined as $\boldsymbol{z} = (\boldsymbol{\theta}, \boldsymbol{p})$, where auxiliary momentum vector $\boldsymbol{p} \sim \mathcal{N}(\mathbf{0}_D, \mathbf{I}_D)$ leads to kinetic energy $\mathcal{G}(\boldsymbol{p}) = \frac{1}{2}\boldsymbol{p}^\mathrm{T}\boldsymbol{p}$, assuming mass matrix $\boldsymbol{M} = \mathbf{I}_D$. Combining potential and kinetic energy, the Hamiltonian is $H(\boldsymbol{z}) = U(\boldsymbol{\theta}) + \frac{1}{2}\boldsymbol{p}^\mathrm{T}\boldsymbol{p}$. This guarantees the augmented stationary distribution $\pi(\boldsymbol{z}) \propto \exp(-H(\boldsymbol{z}))$ aligns with $\pi(\boldsymbol{z}) = p(\boldsymbol{\theta}|\mathcal{D})\phi(\boldsymbol{p})$, where $\phi(\cdot)$ denotes the standard Gaussian PDF.

The dynamic simulation within AM-SGHMC is founded on an Itô diffusion process, expressed by the continuous-time stochastic differential equation:

$$d\boldsymbol{z} = \boldsymbol{f}(\boldsymbol{z})dt + \sqrt{2\boldsymbol{D}(\boldsymbol{z})}d\boldsymbol{W}(t) \qquad (3)$$

where $\boldsymbol{f}(\boldsymbol{z})$, $\boldsymbol{W}(t)$ and $\boldsymbol{D}(\boldsymbol{z})$ are the deterministic drift, Wiener process and

diffusion matrix, respectively. The drift is parameterized as:

$$f(z) = -[D(z) + Q(z)]\nabla_z H(z) + \Gamma(z) \quad (4)$$

$$\Gamma(z) = (\nabla_z \cdot [D(z) + Q(z)]^T)^T \quad (5)$$

where $Q(z)$ and $\Gamma(z)$ represent the curl matrix and a correction term, respectively. AM-SGHMC parameterizes the curl and diffusion matrices using two embedded NNs $f_{\phi_Q}(\cdot,\cdot,\cdot) > 0$ and $f_{\phi_D}(\cdot,\cdot,\cdot,\cdot) > 0$ as:

$$Q(z) = \begin{bmatrix} 0 & -G(z) \\ G(z) & 0 \end{bmatrix} \quad (6)$$

$$D(z) = \begin{bmatrix} 0 & 0 \\ 0 & C(z) \end{bmatrix} \quad (7)$$

where the elements of the gyroscopic-coupling matrix $G(z) = diag[f_{\phi_Q}(z)]$ and damping matrix $C(z) = diag[f_{\phi_D}(z)]$ are:

$$f_{\phi_Q,i}(z) = \sigma_i \cdot \left(c_1 + f_{\phi_Q}(\widehat{U}(\theta), p_i, Cate_i)\right) \quad (8)$$

$$f_{\phi_D,i}(z) = c_2 + f_{\phi_D}(\widehat{U}(\theta), p_i, \partial_{\theta_i} \widehat{U}^*(\theta), Cate_i) \quad (9)$$

where $c_1, c_2 > 0$ prevent matrix degeneracy, and $\widehat{U}(\theta) = \frac{U(\theta) - \mu_U}{\sqrt{2D}\sigma_U}$ normalizes the potential energy with its mean $\mu_U$ and standard deviation $\sigma_U$. The gradient term is scaled as $\partial_{\theta_i} \widehat{U}^*(\theta) = \sigma_i \cdot \partial_{\theta_i} \widehat{U}(\theta)$, while $\sigma_i$, $p_i$ and $Cate_i$ represent the estimated scale, momentum and parameter category for the $i$-th component $\theta_i$ of $\theta$, respectively. For detailed NN architectures of $f_{\phi_Q}$ and $f_{\phi_D}$, refer to [31].

The forms of Eqs. (8)-(9) theoretically guarantee scale- and translation-invariance in the sampling performance relative to the posterior PDF [31]. However, its component-independent form — designed for meta-learning across tasks with varying dimensions — can only accelerate local conditional component-wise sampling. When components are strongly correlated, exploring narrow high-probability paths remains slow, making inter-component correlation the primary bottleneck for sampling efficiency. Moreover, discrepancies in optimal sampling strategies under varying correlation conditions also constrain generalization capability.

The statistics $\sigma_i$, $\mu_U$, and $\sigma_U$ are initially unknown and adaptively estimated during burn-in phase using exponential moving average (EMA), extending the moment estimation process adopted in Adam [34]. The decay rates for the first two moments, $\beta_1$ and $\beta_2$, are determined by the equivalent cumulative time step $\hat{t}$ and the number of chains $K$ as:

$$\beta_1 = \frac{\hat{t} - 1}{\hat{t}} \tag{10}$$

$$\beta_2 = \frac{\hat{t}K - K - 1}{\hat{t}K - 1} \tag{11}$$

Let $y$ represent the variable to be estimated, $\theta_i$ or $U(\boldsymbol{\theta})$. $\{y_{t,k}\}_{k=1}^{K}$ denotes its simulated values across $K$ chains at step $t > 0$. The biased estimates, $m_t$ and $v_t$, of its 1st moment and centered 2nd moment are updated as:

$$m_t = \beta_1 m_{t-1} + (1 - \beta_1)\left[\frac{1}{K}\sum_{k=1}^{K} y_{t,k}\right] \tag{12}$$

$$v_t = \beta_2 v_{t-1} + (1 - \beta_2)\left[\frac{1}{K}\sum_{k=1}^{K}(y_{t,k} - \widehat{m}_t)^2\right]$$
$$+ \left[\beta_2 + (1 - \beta_2)\frac{1}{K}\right](\widehat{m}_t - \widehat{m}_{t-1})^2 \tag{13}$$

with prior initial values $m_0 = 0$ and $v_0 = v_0^*$. The bias-corrected estimates and their corresponding robust estimates, with initial values taken as prior knowledge, are derived and respectively applied to adaptively estimate $\{\mu_U, \sigma_U^2\}$ and parameter variances $\{\sigma_i^2\}_{i=1}^{D}$ required for the main sampling phase [31].

For adaptive estimation, accuracy depends on the exponential decay rate $\beta$,

controlled by a fixed equivalent cumulative time step $\hat{t}$. In the early stage, flexible estimation (requiring smaller $\hat{t}$) is needed to adapt to rapid Markov chain convergence, while as the chains approach the stationary distribution, higher precision (larger $\hat{t}$) are required for scale normalization. AM-SGHMC's constant $\hat{t}$ forces a trade-off, limiting performance in both phases.

Another challenge is ensuring robust adaptive estimation to prevent sampling divergence caused by estimation bias. While delaying estimation reduces risks associated with samples far from the stationary distribution region, these non-converged samples cannot be fully excluded. The practical dilemma lies in balancing early utilization of samples approaching convergence with non-converged sample rejection. A method to neutralize the impact of harmful samples is thus critical.

Using a modified forward Euler discretization with step-size $\eta$, the AM-SGHMC update rule can be derived from Eqs. (3)-(5):

$$\boldsymbol{p}_{t+1} = \bigl(1 - \eta \boldsymbol{C}(\boldsymbol{z}_t)\bigr)\boldsymbol{p}_t - \eta \boldsymbol{G}(\boldsymbol{z}_t)\boldsymbol{\nabla}_{\boldsymbol{\theta}_t} U(\boldsymbol{\theta}_t)$$

$$+\eta\left[\bigl(\boldsymbol{\nabla}_{\boldsymbol{\theta}_t} \cdot \boldsymbol{G}^T(\boldsymbol{z}_t)\bigr)^T + \bigl(\boldsymbol{\nabla}_{\boldsymbol{p}_t} \cdot \boldsymbol{C}^T(\boldsymbol{z}_t)\bigr)^T\right] + \sqrt{2\eta}\boldsymbol{\epsilon}_t \quad (14)$$

$$\boldsymbol{\theta}_{t+1} = \boldsymbol{\theta}_t + \eta \boldsymbol{G}(\hat{\boldsymbol{z}}_t)\boldsymbol{p}_{t+1} - \eta\bigl(\boldsymbol{\nabla}_{\boldsymbol{p}_{t+1}} \cdot \boldsymbol{G}^T(\hat{\boldsymbol{z}}_t)\bigr)^T \quad (15)$$

where $t$ is the discretized time index; $\boldsymbol{z}_t = (\boldsymbol{\theta}_t, \boldsymbol{p}_t)$, $\hat{\boldsymbol{z}}_t = (\boldsymbol{\theta}_t, \boldsymbol{p}_{t+1})$ and $\boldsymbol{\epsilon}_t \sim \mathcal{N}(\boldsymbol{0}_D, \boldsymbol{C}(\boldsymbol{z}_t))$. Given an initial state $\boldsymbol{z}_0 = (\boldsymbol{\theta}_0, \boldsymbol{p}_0)$, parameter samples are generated by iteratively applying Eqs. (14)-(15).

As for network training, gradient backpropagation occurs every $T$ steps. For a cycle starting at $t_0$, after simulating $K$ parallel Markov chains for $T$ steps, a sample set $\{\{\boldsymbol{\theta}_{t_0+s}^k\}_{s=1}^T\}_{k=1}^K$ is obtained. Denoting the estimated PDF of the first $s$ samples as $\bar{q}_s(\cdot|\mathcal{D})$, the loss function is:

$$Loss_{t_0} = \frac{1}{K}\frac{1}{T}\sum_{k=1}^{K}\sum_{s=1}^{T}U(\boldsymbol{\theta}_{t_0+s}^{k}) + \frac{1}{K}\frac{1}{T-M}\sum_{k=1}^{K}\sum_{s=M+1}^{T}\log \bar{q}_s(\boldsymbol{\theta}_{t_0+s}^{k}|\mathcal{D}) \quad (16)$$

Meanwhile, truncated back-propagation through time (BPTT) is applied at each step to respect Markov chain properties (see [31]). This loss approximates the negative Evidence Lower Bound (ELBO). Thus, minimizing $Loss_{t_0}$ (equivalent to maximizing ELBO) reduces Kullback-Leibler divergence $KL[q \parallel \pi]$, accelerating convergence to the stationary distribution $\pi(\boldsymbol{\theta})$. Post-training, AM-SGHMC achieves superior performance and generalizes well to homogeneous sampling tasks.

## 2.2 Fundamental idea of APM-SGHMC

To enable the sampler of our new method, APM-SGHMC, to overcome the limitations in generalization capability after training—that is, to possess both the efficiency of NN-enhanced approaches and the universality akin to classical methods—we focus on the inherent complexity of the posterior distribution. By transforming domain-specific parameters into rotation-invariant PCs, we bypass the constraints imposed by component-wise NN architectures. This approach not only incorporates rotation-invariance but also preserves the scale- and translation-invariance of AM-SGHMC, thereby achieving universal generalization while further enhancing sampling efficiency.

APM-SGHMC incorporates our newly proposed adaptive PC estimation algorithm as a practical implementation approach. Before technical details, this section provides a high-level blueprint of our algorithm's contributions, adaptive PC transformation, emphasizing its conceptual breakthrough in addressing the primary correlation issue that critically impedes sampling efficiency in AM-SGHMC. The practical implementation details and the entire algorithm will be presented in the next two sections, respectively.

In AM-SGHMC, the adaptive variance estimation generalizes the meta-learning sampling method to possess scale- and translation-invariance but is component-wise, ignoring parameter correlations. APM-SGHMC introduces an adaptive PCD estimation technique, thereby exploiting correlation information from a sample covariance matrix to extend the meta-learning sampling method to rotation-invariance with respect to the posterior PDF, as shown in Fig. 1.

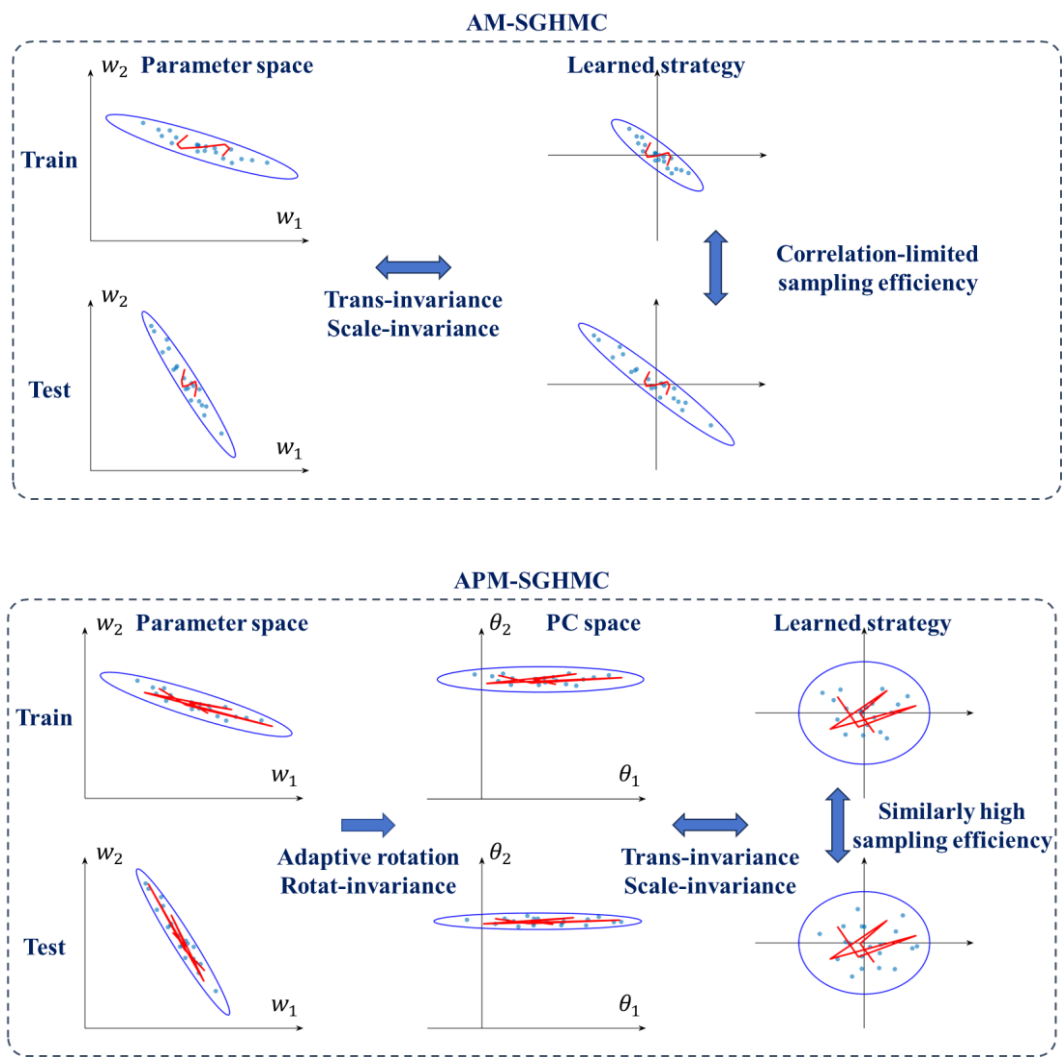

**Figure 1.** Conceptual diagram of the fundamental idea (Red polylines are Markov chains).

Based on the newly proposed adaptive PCD estimation technique, the adaptive estimation process in APM-SGHMC is modified to iteratively perform three sequential steps: first, conducting mean estimation for each parameter identical to the original process; then, executing PCD estimation; and finally, performing variance estimation across these PCs of the parameter vector. This modification upgrades AM-SGHMC's component-wise 2nd-order estimation to a cross-component PC-wise 2nd-order version.

As for the dynamic simulation process, specifically, let $\boldsymbol{P} = (\vec{\boldsymbol{n}}_1, \cdots, \vec{\boldsymbol{n}}_D)$ be the rotation matrix composed of PCDs $\vec{\boldsymbol{n}}_i$, where $i = 1, \cdots, D$. We represent the model parameter vector using the new symbol $\boldsymbol{w}$ and distinguish it from the state $\boldsymbol{\theta}$ of the Markov chain. Then, the relationship between the model parameter vector $\boldsymbol{w}$ and the state $\boldsymbol{\theta}$ can be formulated as a random variable transformation:

$$\boldsymbol{w} = \boldsymbol{P}\boldsymbol{\theta} \tag{17}$$

According to the multivariate change-of-variables theorem in probability theory, under the random variable transformation shown in Eq. (17), the target-distribution PDF of the state $\boldsymbol{\theta}$ is given by:

$$\pi(\boldsymbol{\theta}) = p_{\boldsymbol{\theta}}(\boldsymbol{\theta}) = p_{\boldsymbol{w}}(\boldsymbol{w}) \left| \det\left(\frac{d\boldsymbol{w}}{d\boldsymbol{\theta}}\right) \right| \tag{18}$$

where $p_{\boldsymbol{w}}(\boldsymbol{w})$ is known as the parameter posterior PDF $p_{\boldsymbol{w}}(\boldsymbol{w}) = p(\boldsymbol{w}|\mathcal{D}) \propto p(\mathcal{D}|\boldsymbol{w})p(\boldsymbol{w})$.

Since the rotation matrix $\boldsymbol{P}$ is composed of orthonormal bases $\{\vec{\boldsymbol{n}}_i\}_{i=1}^D$, we know that $\left| \det\left(\frac{d\boldsymbol{w}}{d\boldsymbol{\theta}}\right) \right| = |\det(\boldsymbol{P})| = 1$. Therefore, we have:

$$p_{\boldsymbol{\theta}}(\boldsymbol{\theta}) = p_{\boldsymbol{w}}(\boldsymbol{w}) = p_{\boldsymbol{w}}(\boldsymbol{P}\boldsymbol{\theta}) \tag{19}$$

Furthermore, the potential energy field can be defined as:

$$U_{\boldsymbol{\theta}}(\boldsymbol{\theta}) = U_{\boldsymbol{w}}(\boldsymbol{w}) = U_{\boldsymbol{w}}(\boldsymbol{P}\boldsymbol{\theta}) \tag{20}$$

where $U_{\boldsymbol{w}}(\boldsymbol{w}) = -\log(p(\mathcal{D}|\boldsymbol{w})) - \log(p(\boldsymbol{w})) + c^*$ is known as the potential energy field when no rotation transformation is applied. Then, based on the differential relationship $(\nabla_{\boldsymbol{\theta}} U_{\boldsymbol{\theta}}(\boldsymbol{\theta}))^T d\boldsymbol{\theta} = dU_{\boldsymbol{\theta}}(\boldsymbol{\theta}) = dU_{\boldsymbol{w}}(\boldsymbol{w}) = (\nabla_{\boldsymbol{w}} U_{\boldsymbol{w}}(\boldsymbol{w}))^T d\boldsymbol{w}$, the gradient

of the potential energy field under the rotation transformation can be derived as:

$$\nabla_\theta U_\theta(\theta) = \left(\frac{d\mathbf{w}}{d\theta}\right)^T \nabla_w U_w(\mathbf{w}) = \mathbf{P}^T \nabla_w U_w(\mathbf{P}\theta) \tag{21}$$

In practical applications involving structural dynamics models, compared to the calculations of $U_w(\cdot)$ and $\nabla_w U_w(\cdot)$ required in AM-SGHMC, the additional time consumed by the three matrix multiplications in calculating $U_\theta(\cdot)$ and $\nabla_\theta U_\theta(\cdot)$ in Eqs. (20)-(21) is negligible.

Since $\theta$ represents the components of $\mathbf{w}$ along the PCDs, and regardless of the rotation applied, the relationships among the PCDs are kept invariant (the cases where some PC eigenvalues are equal can be disregarded, given their relatively minor impact on the sampling algorithm), rotation-invariance is exhibited by APM-SGHMC. This property of rotation-invariance further enhances the generalization capability of the NN embedded within the sampling strategy.

It should be noted that each PCD no longer corresponds to a domain-specific parameter category. Given that, compared to highlighting the differences in sampling strategies demanded by different parameter categories, mitigating the correlations among components leads to a more significant improvement in sampling efficiency. Consequently, the APM-SGHMC algorithm discards the parameter category $Cate_i$ in the NN input.

An additional advantage of discarding the parameter category $Cate_i$ is that after being trained on a task involving one model class, the APM-SGHMC sampler has the potential to be directly applied to another task for a significantly different model class. Such an application does not necessitate the two model classes to have parameter categories of the same kind. This universal generalization potential has been verified in the cases of Bayesian inference of structural dynamic models in Section 5.2.

## 3. MAIN CONTRIBUTIONS IN APM-SGHMC

### *3.1 Adaptive estimation technique for PCD*

The challenge in implementing the PC transformation lies in the fact that it is often impossible to obtain a sufficient number of effective samples for statistical purposes. The adaptive PCD estimation technique derived in this section enables finite Markov chains to update the PCD in real-time during the sampling process. This, in turn, allows for the real-time enhancement of sampling efficiency through adaptive PC transformation. The two aspects mutually reinforce each other, ultimately leading to the rapid acquisition of enough effective samples and accurate statistical results.

For a set of $N$ parameter samples $\{w^{(i)}\}_{i=1}^{N}$, and given an arbitrary unit direction vector $\vec{n}$, to achieve the adaptive estimation of the PCDs in PC analysis (PCA) [35-36], a statistical vector is first constructed as follows:

$$\vec{p}(\vec{n}) = \frac{1}{N}\sum_{i=1}^{N}\langle w^{(i)}, \vec{n}\rangle w^{(i)} \tag{22}$$

where $\langle a, b\rangle = a^T \cdot b$ represents the inner product of any two vectors $a$ and $b$. Moreover, in this section, it is assumed that the mean of the sample set $\{w^{(i)}\}_{i=1}^{N}$ is zero, which can be achieved by mean-centering preprocessing. To ensure distinction from symbols such as the momentum $p$ and the PDF $p$, arrow annotations are used on the column vectors related to PCDs in this study.

The direction of the statistical vector $\vec{p}(\vec{n})$ is denoted as:

$$\vec{n}_{\vec{p}}(\vec{n}) = \frac{\vec{p}(\vec{n})}{|\vec{p}(\vec{n})|} \tag{23}$$

In fact, Appendix A demonstrates that Eqs. (22)-(23) are equivalent to a single iteration in the eigen-decomposition of the covariance matrix of the sample set $\{w^{(i)}\}_{i=1}^{N}$ using the power iteration method. Consequently, the direction $\vec{n}_{\vec{p}}(\vec{n})$ has a higher alignment with the 1st PC subspace $S_1$ than the initial direction $\vec{n}$. This relationship is quantified by the following inequality:

$$\langle \vec{n}_{\vec{p}}(\vec{n}), \vec{n}_1 \rangle^2 \geq \langle \vec{n}, \vec{n}_1 \rangle^2 \tag{24}$$

where $\vec{n}_1$ is any unit vector in $S_1$. The equality holds if and only if one of the following two conditions is satisfied: (1) $\vec{n}$ already lies in $S_1$, i.e., $\vec{n} \in S_1$ or (2) $\vec{n}$ is orthogonal to $S_1$, i.e., $\langle \vec{n}, \vec{n}_1 \rangle = 0$ for all $\vec{n}_1 \in S_1$. This implies that, excluding the case where $\vec{n}$ is orthogonal to $S_1$, for any initial direction $\vec{n}$, we can obtain a 1st PCD $\vec{n}_1 \in S_1$ by repeatedly using the new direction $\vec{n}_{\vec{p}}$ as the input for the statistical vector ($\vec{n} \leftarrow \vec{n}_{\vec{p}}$) and iterating according to Eqs. (22)-(23).

For all $D$ PCs in a $D$-dimensional parameter space, given the first $d-1$ obtained PCDs $\{\vec{n}_{\vec{p}_j}\}_{j=1}^{d-1}$, the residual components of each sample can be computed as:

$$w^{(i,d)} = w^{(i)} - \sum_{j=1}^{d-1} \langle w^{(i)}, \vec{n}_{\vec{p}_j} \rangle \vec{n}_{\vec{p}_j} \tag{25}$$

These residuals are then used to iteratively obtain the $d$-th PCD:

$$\vec{p}_d \leftarrow \frac{1}{N} \sum_{i=1}^{N} \langle w^{(i,d)}, \vec{n}_{\vec{p}_d} \rangle w^{(i,d)} \tag{26}$$

$$\vec{n}_{\vec{p}_d} = \frac{\vec{p}_d}{|\vec{p}_d|} \tag{27}$$

Furthermore, to achieve adaptive estimation of PCDs, analogous to adaptive estimation of the 1st moment in Eq. (12), the arithmetic average in Eq. (26) is reformulated as an EMA process:

$$\vec{p}_d \leftarrow \beta_3 \vec{p}_d + (1 - \beta_3) \langle w^{(i,d)}, \vec{n}_{\vec{p}_d} \rangle w^{(i,d)} \tag{28}$$

Here, the exponential decay rate $\beta_3$ is determined by the equivalent cumulative time step $\hat{t}_3$:

$$\beta_3 = \frac{\hat{t}_3 - 1}{\hat{t}_3} \tag{29}$$

In the APM-SGHMC algorithm, $K$ parallel Markov chains are simultaneously simulated for $T_{Ada}$ steps during the adaptive estimation phase, continuously generating parameter samples $\{w^{(t,k)}\}_{k=1}^{K}$ for $t = 1, \cdots, T_{Ada}$. For the $K$ parameter samples $\{w^{(t,k)}\}_{k=1}^{K}$ obtained at step $t$, Eq. (28) is reformulated as:

$$\vec{p}_{d,t} = \beta_3 \vec{p}_{d,t-1} + (1 - \beta_3) \frac{1}{K} \sum_{k=1}^{K} \langle w^{(t,k,d)}, \vec{n}_{\vec{p}_d, t-1} \rangle w^{(t,k,d)} \tag{30}$$

To maintain orthogonality among updated PCDs, when $d \neq 1$, the vectors $\vec{p}_{d,t-1}$ and $\vec{n}_{\vec{p}_d,t-1}$ are orthogonalized prior to substitution into Eq. (30) as:

$$\vec{p}_{d,t-1} \leftarrow \frac{\vec{p}_{d,t-1} - \sum_{j=1}^{d-1} \langle \vec{p}_{d,t-1}, \vec{n}_{\vec{p}_j,t} \rangle \vec{n}_{\vec{p}_j,t}}{\left| \vec{n}_{\vec{p}_d,t-1} - \sum_{j=1}^{d-1} \langle \vec{n}_{\vec{p}_d,t-1}, \vec{n}_{\vec{p}_j,t} \rangle \vec{n}_{\vec{p}_j,t} \right|} \tag{31}$$

$$\vec{n}_{\vec{p}_d,t-1} \leftarrow \frac{\vec{n}_{\vec{p}_d,t-1} - \sum_{j=1}^{d-1} \langle \vec{n}_{\vec{p}_d,t-1}, \vec{n}_{\vec{p}_j,t} \rangle \vec{n}_{\vec{p}_j,t}}{\left| \vec{n}_{\vec{p}_d,t-1} - \sum_{j=1}^{d-1} \langle \vec{n}_{\vec{p}_d,t-1}, \vec{n}_{\vec{p}_j,t} \rangle \vec{n}_{\vec{p}_j,t} \right|} \tag{32}$$

After updating all $D$ PCDs, the estimated rotation matrix at step $t$ is constructed as:

$$P_t = (\vec{n}_{\vec{p}_1,t}, \cdots, \vec{n}_{\vec{p}_D,t}) \tag{33}$$

Finally, recall the special case excluded during the discussion of Eq. (24): $\vec{n}$ is orthogonal to $S_1$. Since all estimated PCDs $\{\vec{n}_{\vec{p}_d}\}_{d=1}^{D}$ span the complete $D$-dimensional parameter space, there must exist at least one estimated PCD $\vec{n}_{\vec{p}_d}$ with a non-zero projection onto $S_1$, i.e., there exists $\vec{n}_1 \in S_1$ such that $\langle \vec{n}_{\vec{p}_d}, \vec{n}_1 \rangle \neq 0$.

Consequently, when the estimated 1st PCD $\vec{n}_{\vec{p}_1}$ is orthogonal to subspace $S_1$, the sample variance $\sigma_1^2$ along $\vec{n}_{\vec{p}_1}$ will eventually be surpassed by $\sigma_d^2$ (where $d \neq 1$) as iterations proceed. Given that the standard deviations $\{\sigma_i\}_{i=1}^{D}$ of all PCs are also adaptively estimated in the APM-SGHMC algorithm, this issue can be resolved by rearranging the PCs based on their estimated standard deviations when a significant violation of descending order occurs. Specifically, when the condition $\vee_{d=1}^{D-1}(\sigma_{d+1} > 1.1\sigma_d)$ holds true, the PCs are rearranged to enforce a descending standard deviation sequence.

In fact, due to the stochasticity of parameter samples in the adaptive estimation process, the subspace $S_1$ will evolve over time. When the iteration duration is sufficiently long, even without rearranging, the estimated 1st PCD $\vec{n}_{\vec{p}_1}$ will eventually converge to lie within $S_1$. However, the rearranging operation is retained to facilitate the convergence of the adaptive estimation process. Based on the above description, the adaptive estimation technique for PCD is summarized in Algorithm 1.

---

**Algorithm 1:** Adaptive PCD estimate.

---

**Require:** $\beta_3 \in [0,1)$: Exponential decay rate for PCD estimate
**Require:** $K$: Batch size of samples obtained at each timestep
**Require:** $\boldsymbol{\mu}_{w,1}, \cdots, \boldsymbol{\mu}_{w,T}$: Mean vectors estimated at subsequent timesteps $t = 1, \cdots, T_{Ada}$
**Require:** $\{\sigma_{i,0}\}_{i=1}^{D}, \cdots, \{\sigma_{i,T-1}\}_{i=1}^{D}$: Standard deviations estimated at subsequent timesteps $t = 1, \cdots, T_{Ada}$
**Require:** $\{\boldsymbol{\theta}^{(1,k)}\}_{k=1}^{K}, \cdots, \{\boldsymbol{\theta}^{(T,k)}\}_{k=1}^{K}$: Samples obtained at subsequent timesteps $t = 1, \cdots, T_{Ada}$

  $\boldsymbol{P}_0 = (\vec{n}_{\vec{p}_1,0}, \cdots, \vec{n}_{\vec{p}_D,0}) \leftarrow \boldsymbol{I}_D$ (Initialize PCD vectors)
  **for** $d = 1, \cdots, D$ **do**
    $\vec{p}_{d,0} \leftarrow \sigma_{d,0} \vec{n}_{\vec{p}_d,0}$ (Initialize PCD statistical vectors for each dimension)
  **end for**
  $t \leftarrow 0$ (Initialize timestep)
  **for** $t' = 1, \cdots, T_{Ada}$ **do**
    $t \leftarrow t + 1$
    **if** $\vee_{d=1}^{D-1}(\sigma_{d+1,t-1} > 1.1\sigma_{d,t-1})$ **do**
      Rearrange PCs in descending order of $\{\sigma_{i,t-1}\}_{i=1}^{D}$
      Correspondingly update $\{\sigma_{i,t-1}\}_{i=1}^{D}$, $\{\boldsymbol{\theta}^{(t,k)}\}_{k=1}^{K}$, $\boldsymbol{P}_t$ and $\{\vec{p}_{d,t}\}_{d=1}^{D}$
    **end if**

---

$$\{w^{(t,k)}\}_{k=1}^{K} \leftarrow \{P_{t-1}\theta^{(t,k)} - \mu_{w,t}\}_{k=1}^{K} \text{(Compute centered parameter samples)}$$

for $d = 1, \cdots, D$ do

  if $d = 1$ do

    $w^{(t,k,d)} \leftarrow w^{(t,k)}$

  else do (Orthogonalize against previous dimensions)

$$w^{(t,k,d)} \leftarrow w^{(t,k)} - \sum_{j=1}^{d-1} \langle w^{(t,k)}, \vec{n}_{\vec{p}_j,t} \rangle \vec{n}_{\vec{p}_j,t}$$

$$\vec{p}_{d,t-1} \leftarrow \frac{\vec{p}_{d,t-1} - \sum_{j=1}^{d-1}\langle \vec{p}_{d,t-1}, \vec{n}_{\vec{p}_j,t}\rangle \vec{n}_{\vec{p}_j,t}}{\left|\vec{n}_{\vec{p}_d,t-1} - \sum_{j=1}^{d-1}\langle \vec{n}_{\vec{p}_d,t-1},\vec{n}_{\vec{p}_j,t}\rangle \vec{n}_{\vec{p}_j,t}\right|}$$

$$\vec{n}_{\vec{p}_d,t-1} \leftarrow \frac{\vec{n}_{\vec{p}_d,t-1} - \sum_{j=1}^{d-1}\langle \vec{n}_{\vec{p}_d,t-1},\vec{n}_{\vec{p}_j,t}\rangle \vec{n}_{\vec{p}_j,t}}{\left|\vec{n}_{\vec{p}_d,t-1} - \sum_{j=1}^{d-1}\langle \vec{n}_{\vec{p}_d,t-1},\vec{n}_{\vec{p}_j,t}\rangle \vec{n}_{\vec{p}_j,t}\right|}$$

  end if

$$\vec{p}_{d,t}^{ipt} \leftarrow \frac{1}{K}\sum_{k=1}^{K}\langle w^{(t,k,d)}, \vec{n}_{\vec{p}_d,t-1}\rangle w^{(t,k,d)}$$

$\vec{p}_{d,t} \leftarrow \beta_3 \vec{p}_{d,t-1} + (1-\beta_3)\vec{p}_{d,t}^{ipt}$ (Update PCD statistical vector)

$\vec{n}_{\vec{p}_d,t} = \frac{\vec{p}_{d,t}}{|\vec{p}_{d,t}|}$ (Normalize to get PCD vector)

end for

$P_t \leftarrow (\vec{n}_{\vec{p}_1,t}, \cdots, \vec{n}_{\vec{p}_D,t})$ (Construct rotation matrix)

$\{\theta^{(t,k)}\}_{k=1}^{K} \leftarrow \{P_t^T P_{t-1}\theta^{(t,k)}\}_{k=1}^{K}$ (Update state samples)

end for

**return** $\{\theta^{(t,k)}\}_{k=1}^{K}$ at timesteps $t = 1, \cdots, T_{Ada}$ (Updating state samples)

**return** $P_t$ at timesteps $t = 1, \cdots, T_{Ada}$ (Estimating PCD rotation matrix)

Additionally, Bayesian model classes often incorporate a predictive error parameter $\sigma$ that exhibits nonlinear dependency on other structural model parameters, yet remains locally linearly independent in the vicinity of the optimal solution. Formally, this occurs because $\sigma$ attains its minimum at the optimal point $w^*$, leading to $\nabla_w \sigma(w^*) = \mathbf{0}_D$. Consequently, we recommend that within the adaptive PCD estimation framework, $\sigma$ should not participate in rotational transformations to preserve the statistical properties of the Bayesian model class.

### *3.2 Adaptive estimation with decay rates evolving over iterations*

Although for the purpose of enhancing estimation accuracy, the later-stage estimated values need to be stable, which implies a larger exponential decay rate $\beta$, the rapid convergence of early-stage samples towards high-probability regions and the swift directional rotations brought about by PCD estimation both demand greater flexibility in the estimated values, which means a smaller $\beta$. Consequently, setting $\beta$ as a constant no longer meets the computational requirements.

The exponential decay rates in Eqs. (10)-(11) are first decoupled and then generalized, together with the exponential decay rate in Eq. (29), to rates that vary with the time step $t$:

$$\beta_{i,t} = \frac{\hat{t}_{i,t} - 1}{\hat{t}_{i,t}}, i = 1, 3, t = 1, \cdots, T_{Ada} \tag{34}$$

$$\beta_{2,t} = \frac{\hat{t}_{2,t} K - K - 1}{\hat{t}_{2,t} K - 1}, t = 1, \cdots, T_{Ada} \tag{35}$$

where $\{\hat{t}_{i,t}\}_{t=1}^{T_{Ada}}, i = 1, 2, 3$ represent three sets of equivalent cumulative time steps that vary with the time step $t$.

Correspondingly, the EMA process is rewritten as:

$$\boldsymbol{m}_t = \beta_{1,t} \boldsymbol{m}_{t-1} + (1 - \beta_{1,t}) \left[ \frac{1}{K} \sum_{k=1}^{K} \boldsymbol{w}^{(t,k)} \right] \tag{36}$$

$$\vec{\boldsymbol{p}}_{d,t} = \beta_{3,t} \vec{\boldsymbol{p}}_{d,t-1} + (1 - \beta_{3,t}) \frac{1}{K} \sum_{k=1}^{K} \langle \boldsymbol{w}^{(t,k,d)}, \vec{\boldsymbol{n}}_{\vec{p}_{d,t-1}} \rangle \boldsymbol{w}^{(t,k,d)} \tag{37}$$

$$\boldsymbol{v}_t = \beta_{2,t} \boldsymbol{v}_{t-1} + (1 - \beta_{2,t}) \left[ \frac{1}{K} \sum_{k=1}^{K} (\boldsymbol{\theta}^{(t,k)} - \boldsymbol{P}_t^T \widehat{\boldsymbol{m}}_t)^2 \right]$$

$$+ (1 - \beta_{2,t}) \left[ \frac{\beta_{2,t}}{1 - \beta_{2,t}} + \frac{1}{K} \right] \frac{\hat{t}_{1,t}(\hat{t}_{1,t} - 1)}{\hat{t}_{2,t}(\hat{t}_{2,t} - 1)} (\boldsymbol{P}_t^T \widehat{\boldsymbol{m}}_t - \boldsymbol{P}_t^T \widehat{\boldsymbol{m}}_{t-1})^2 \tag{38}$$

where the squaring operations of vectors are elementwise, and the factor $\frac{\hat{t}_{1,t}(\hat{t}_{1,t}-1)}{\hat{t}_{2,t}(\hat{t}_{2,t}-1)}$ is a correction coefficient used to mitigate the effects of decoupling $\beta_{1,t}$ and $\beta_{2,t}$. Furthermore, the bias-corrected estimates, $\widehat{\boldsymbol{m}}_t$ and $\widehat{\boldsymbol{v}}_t$, can be computed as:

$$\widehat{m}_t = \frac{m_t}{1 - \prod_{i=1}^{t} \beta_{1,i}} \tag{39}$$

$$\widehat{v}_t = \frac{v_t - (\prod_{i=1}^{t} \beta_{2,i})v_0^*}{1 - \prod_{i=1}^{t} \beta_{2,i}} \tag{40}$$

A simple proof of Eqs. (39)-(40) is provided in Appendix B. Then, the corresponding robust estimates, taking initial values $\mathbf{0}$ and $v_0^*$ as prior knowledge, are modified as:

$$\widehat{m}_t = (1 + \prod_{i=1}^{t} \beta_{1,i})m_t \tag{41}$$

$$\widehat{v}_t = v_t + (\prod_{i=1}^{t} \beta_{2,i})(v_t - v_0^*) \tag{42}$$

which will be biased towards the prior initial values $\mathbf{0}$ and $v_0^*$ during the initial timesteps.

In the APM-SGHMC algorithm, to meet the algorithm's requirements while facilitating computation, we recommend setting the sequence of equivalent cumulative time steps as follows:

$$\hat{t}_{i,t} = \hat{t}_{i,t}(t; t_i^b, \hat{t}_{i,min}, \hat{t}_{i,max}) = \begin{cases} \hat{t}_{i,min}, & t - t_i^b < \hat{t}_{i,min} \\ t - t_i^b, & \hat{t}_{i,min} \leq t - t_i^b \leq \hat{t}_{i,max} \\ \hat{t}_{i,max}, & t - t_i^b > \hat{t}_{i,max} \end{cases} \tag{43}$$

Here, the early-stage equivalent cumulative time steps $\hat{t}_{i,min}$ enable flexible adjustments during the early adaptive phase, while the late-stage equivalent cumulative time steps $\hat{t}_{i,max} > \hat{t}_{i,min}$ ensure the final estimation accuracy in the later stable phase. We recommend setting a waiting period of at least $t_i^b \geq 0.1\hat{t}_{i,min}$ steps during the initial phase, so that the adaptive estimation process consistently approaches the ideal accuracy controlled by $\hat{t}_{i,t}$, as analyzed in Appendix C.

Based on the improvements presented in this section, the complete adaptive PC estimation algorithm within APM-SGHMC is provided in Algorithm 2.

**Algorithm 2:** Adaptive PC estimate of the parameter vector in APM-SGHMC.

**Require:** $t_i^b, \hat{t}_{i,min}, \hat{t}_{i,max}, i = 1,2,3$: Exponential decay parameters for adaptive estimates

**Require:** $K$: Batch size of samples obtained at each timestep

**Require:** $\mathbf{v}_0^* = (\sigma_{1,0}^2, \cdots, \sigma_{D,0}^2)^T$: Prior initial value vector of the variances

**Require:** $\{\boldsymbol{\theta}^{(1,k)}\}_{k=1}^K, \cdots, \{\boldsymbol{\theta}^{(T_{Ada},k)}\}_{k=1}^K$: Samples obtained at subsequent timesteps $t = 1, \cdots, T_{Ada}$

  for $i = 1, 2, 3$ do
    $\hat{t}_{i,0} \leftarrow \hat{t}_{i,min}$
    Compute $\beta_{i,0}$ as in Eqs. (34)-(35)
  end for
  $\mathbf{m}_0 \leftarrow \mathbf{0}$ (Initialize 1st moment vector)
  $\mathbf{P}_0 = (\vec{\mathbf{n}}_{\bar{p}_1,0}, \cdots, \vec{\mathbf{n}}_{\bar{p}_D,0}) \leftarrow \mathbf{I}_D$ (Initialize PCD vectors)
  for $d = 1, \cdots, D$ do
    $\vec{\mathbf{p}}_{d,0} \leftarrow \sigma_{d,0}^2 \vec{\mathbf{n}}_{\bar{p}_d,0}$ (Initialize PCD statistical vectors for each dimension)
  end for
  $\mathbf{v}_0 \leftarrow \mathbf{v}_0^*$ (Initialize variance vector)
  $\Pi_{1,0} \leftarrow 1$ (Initialize product of $\beta_{1,t}$)
  $\Pi_{2,0} \leftarrow 1$ (Initialize product of $\beta_{2,t}$)
  if $\vee_{d=1}^{D-1}(\sigma_{d+1,0} > 1.1\sigma_{d,0})$ do
    Rearrange PCs in descending order of $\{\sigma_{i,0}\}_{i=1}^D$
    Correspondingly update $\{\sigma_{i,0}\}_{i=1}^D$, $\mathbf{P}_0$ and $\{\vec{\mathbf{p}}_{d,0}\}_{d=1}^D$
  end if
  $t \leftarrow 0$ (Initialize timestep)
  for $t' = 1, \cdots, T_{Ada}$ do
    $t \leftarrow t + 1$
    for $i = 1, 2, 3$ do
      if $\hat{t}_{i,min} \leq t - t_i^b \leq \hat{t}_{i,max}$ do
        $\hat{t}_{i,t} \leftarrow \hat{t}_{i,t-1} + 1$
        Compute $\beta_{i,t}$ as in Eqs. (34)-(35)
      else do
        $\hat{t}_{i,t} \leftarrow \hat{t}_{i,t-1}$
        $\beta_{i,t} \leftarrow \beta_{i,t-1}$
      end if
    end for
    $\Pi_{1,t} \leftarrow \beta_{1,t}\Pi_{1,t-1}$ (Update product of $\beta_{1,t}$)
    $\Pi_{2,t} \leftarrow \beta_{2,t}\Pi_{2,t-1}$ (Update product of $\beta_{2,t}$)
    $\{\mathbf{w}^{(t,k)}\}_{k=1}^K \leftarrow \{\mathbf{P}_{t-1}\boldsymbol{\theta}^{(t,k)}\}_{k=1}^K$ (Compute parameter samples)
    $\overline{\mathbf{w}_t} \leftarrow \frac{1}{K}\sum_{k=1}^K \mathbf{w}^{(t,k)}$
    $\mathbf{m}_t \leftarrow \beta_{1,t}\mathbf{m}_{t-1} + (1 - \beta_{1,t})\overline{\mathbf{w}_t}$ (Update biased 1st moment estimate)

$\widehat{m}_t \leftarrow m_t * (1 + \Pi_{1,t})$ (Compute robust bias-corrected 1st moment estimate)

$\{w^{(t,k)}\}_{k=1}^{K} \leftarrow \{w^{(t,k)} - \widehat{m}_t\}_{k=1}^{K}$ (Compute centered parameter samples)

**for** $d = 1, \cdots, D$ **do**

    **if** $d = 1$ **do**

        $w^{(t,k,d)} \leftarrow w^{(t,k)}$

    **else do** (Orthogonalize against previous dimensions)

        $w^{(t,k,d)} \leftarrow w^{(t,k)} - \sum_{j=1}^{d-1} \langle w^{(t,k)}, \vec{n}_{\vec{p}_{j,t}} \rangle \vec{n}_{\vec{p}_{j,t}}$

        $\vec{p}_{d,t-1} \leftarrow \frac{\vec{p}_{d,t-1} - \sum_{j=1}^{d-1} \langle \vec{p}_{d,t-1}, \vec{n}_{\vec{p}_{j,t}} \rangle \vec{n}_{\vec{p}_{j,t}}}{\left| \vec{n}_{\vec{p}_{d,t-1}} - \sum_{j=1}^{d-1} \langle \vec{n}_{\vec{p}_{d,t-1}}, \vec{n}_{\vec{p}_{j,t}} \rangle \vec{n}_{\vec{p}_{j,t}} \right|}$

        $\vec{n}_{\vec{p}_{d,t-1}} \leftarrow \frac{\vec{n}_{\vec{p}_{d,t-1}} - \sum_{j=1}^{d-1} \langle \vec{n}_{\vec{p}_{d,t-1}}, \vec{n}_{\vec{p}_{j,t}} \rangle \vec{n}_{\vec{p}_{j,t}}}{\left| \vec{n}_{\vec{p}_{d,t-1}} - \sum_{j=1}^{d-1} \langle \vec{n}_{\vec{p}_{d,t-1}}, \vec{n}_{\vec{p}_{j,t}} \rangle \vec{n}_{\vec{p}_{j,t}} \right|}$

    **end if**

    $\vec{p}_{d,t}^{ipt} \leftarrow \frac{1}{K} \sum_{k=1}^{K} \langle w^{(t,k,d)}, \vec{n}_{\vec{p}_{d,t-1}} \rangle w^{(t,k,d)}$

    $\vec{p}_{d,t} \leftarrow \beta_{3,t} \vec{p}_{d,t-1} + (1 - \beta_{3,t}) \vec{p}_{d,t}^{ipt}$ (Update PCD statistical vector)

    $\vec{n}_{\vec{p}_{d,t}} = \frac{\vec{p}_{d,t}}{|\vec{p}_{d,t}|}$ (Normalize to get PCD vector)

**end for**

$P_t \leftarrow (\vec{n}_{\vec{p}_{1,t}}, \cdots, \vec{n}_{\vec{p}_{D,t}})$ (Construct rotation matrix)

$\{\theta^{(t,k)}\}_{k=1}^{K} \leftarrow \{P_t^T P_{t-1} \theta^{(t,k)}\}_{k=1}^{K}$ (Update state samples)

**if** $t = 1$ **do**

    $\widehat{m}_0 \leftarrow \widehat{m}_1$

**end if**

$v_t^{ipt} \leftarrow \frac{1}{K} \sum_{k=1}^{K} (\theta^{(t,k)} - P_t^T \widehat{m}_t)^2 + \left( \frac{\beta_{2,t}}{1-\beta_{2,t}} + \frac{1}{K} \right) \frac{\hat{t}_{1,t}(\hat{t}_{1,t}-1)}{\hat{t}_{2,t}(\hat{t}_{2,t}-1)} (P_t^T \widehat{m}_t - P_t^T \widehat{m}_{t-1})^2$

$v_t \leftarrow \beta_{2,t} v_{t-1} + (1 - \beta_{2,t}) v_t^{ipt}$ (Update biased variance estimate)

$\widehat{v}_t = (\sigma_{1,t}^2, \cdots, \sigma_{D,t}^2)^T \leftarrow v_t + \Pi_{2,t}(v_t - v_0^*)$ (Compute robust bias-corrected variance estimate)

**if** $\bigvee_{d=1}^{D-1} (\sigma_{d+1,t} > 1.1 \sigma_{d,t})$ **do**

    Rearrange PCs in descending order of $\{\sigma_{i,t}\}_{i=1}^{D}$

    Correspondingly update $\{\sigma_{i,t}\}_{i=1}^{D}$, $\{\theta^{(t,k)}\}_{k=1}^{K}$, $P_t$ and $\{\vec{p}_{d,t}\}_{d=1}^{D}$

**end if**

**end for**

**return** $\{\theta^{(t,k)}\}_{k=1}^{K}$ at timesteps $t = 1, \cdots, T_{Ada}$ (Updating state samples)

**return** $P_t$ at timesteps $t = 1, \cdots, T_{Ada}$ (Estimating the PCD vectors)

**return** $\widehat{v}_t$ at timesteps $t = 1, \cdots, T_{Ada}$ (Estimating the variances)

## 3.3 Convergence diagnostic and divergence mitigation based on potential energy

In the adaptive estimation process, the relatively small equivalent cumulative time steps $\hat{t}_{i,min}$ set in the early stages make the estimates more susceptible to perturbations from outliers induced by non-converged samples. Excessively large scale-estimates can even cause the Markov chain to diverge. Initially, during the first $T_s$ steps, we scale down the adaptive scale estimates as $\{\sigma_{d,t}\}_{d=1}^{D} \leftarrow \{0.5\sigma_{d,t}\}_{d=1}^{D}$ to mitigate the impact when the estimates are overly large. Simultaneously, we reduce the energy input as $\epsilon_t \leftarrow 0.5\epsilon_t$ to the simulated particle system, enabling the Markov chain to stably approach high-probability regions in a manner more akin to an optimization algorithm. An additional benefit of this treatment is a higher probability of approaching the maximum probability point, thereby obtaining more accurate sample potential energy minima, denoted as $\min\{U\}$, the utility of which will be seen in this section. However, these treatments only provide superficial relief, and outliers can still affect the stability of the estimates. Therefore, a technique for convergence diagnostic and divergence mitigation is required.

It is observed that when the parameter distribution undergoes an affine transformation, the potential energy of all samples that are covariant with this distribution will shift by a same constant value, denoted as $U'(\boldsymbol{\theta}') = U(\boldsymbol{\theta}) + c$. This implies that the discrepancy between the mean of the sample potential energy, $\mu_U$, and the minimum value $\min\{U|c\}$, along with the variance of the sample potential energy, $\sigma_U^2$, remain invariant under such transformations. Given that any non-degenerate Gaussian distribution can be transformed into a standard Gaussian distribution via an affine transformation, we can leverage these affine invariant statistics associated with the standard Gaussian distribution as a reference for calculating those of any arbitrary Gaussian distribution.

It can be readily calculated that the mean of the sample potential energy $U_G$ for a $D$-dimensional Gaussian distribution is given by:

$$\mu_{U_G} = \frac{D}{2} + min\{U_G|c\} \tag{44}$$

and the variance of $U_G$ is:

$$\sigma_{U_G}^2 = \frac{D}{2} \tag{45}$$

One immediate benefit is that this provides a reliable initial value for the adaptive estimation of the variance of the sample potential energy in adaptive sampling algorithms, specifically $\sigma_{U,0}^2 = \frac{D}{2}$. This allows us to set a larger exponential decay rate $\beta_2$, rendering the estimate of $\sigma_U^2$ more stable and precise. Consequently, with $\sigma_{U,0}^2$ taken as prior knowledge, its bias-corrected estimate can be revised as the robust version:

$$\sigma_{U,t}^2 = \hat{v}_t = v_t + \beta_2^t \left(v_t - \frac{D}{2}\right) \tag{46}$$

Thus, the adaptive moment estimation algorithm for the potential energy $U(\boldsymbol{\theta})$ can be refined into Algorithm 3.

---

**Algorithm 3:** Adaptive moment estimate of the potential energy in APM-SGHMC.

**Require:** $\beta_1, \beta_2 \in [0,1)$: Exponential decay rates for adaptive estimates
**Require:** $K$: Batch size of samples obtained at each timestep
**Require:** $D$: Dimension of the state space
**Require:** $\{y_{1,k}\}_{k=1}^K, \cdots, \{y_{T_0+T_{Ada},k}\}_{k=1}^K$: Samples of $U(\boldsymbol{\theta})$ obtained at subsequent timesteps $t = 1, \cdots, T_0 + T_{Ada}$

$m_0 \leftarrow 0$ (Initialize 1st moment estimate)

$v_0 \leftarrow \frac{D}{2}$ (Initialize variance estimate)

$t \leftarrow 0$ (Initialize timestep)

$\hat{t}_1 \leftarrow \frac{1}{1-\beta_1}$ (Compute the equivalent cumulative time step of 1st moment estimate)

$\hat{t}_2 \leftarrow \frac{1}{1-\beta_2} + \frac{1}{K}$ (Compute the equivalent cumulative time step of variance estimate)

**for** $t' = 1, \cdots, T_0 + T_{Ada}$ **do**

  $t \leftarrow t + 1$

  $\bar{y}_t \leftarrow \frac{1}{K}\sum_{k=1}^{K} y_{t,k}$

  $m_t \leftarrow \beta_1 m_{t-1} + (1-\beta_1)\bar{y}_t$ (Update biased 1st moment estimate)

  $\hat{m}_t = \mu_{U,t} \leftarrow m_t/(1-\beta_1^t)$ (Compute bias-corrected 1st moment estimate)

  **if** $t = 1$ **do**

    $\hat{m}_0 \leftarrow \hat{m}_1$

  **end if**

  $v_t^{ipt} \leftarrow \frac{1}{K}\sum_{k=1}^{K}(y_{t,k} - \hat{m}_t)^2 + \left(\frac{\beta_2}{1-\beta_2} + \frac{1}{K}\right)\frac{\hat{t}_1(\hat{t}_1-1)}{\hat{t}_2(\hat{t}_2-1)}(\hat{m}_t - \hat{m}_{t-1})^2$

  $v_t \leftarrow \beta_2 v_{t-1} + (1-\beta_2)v_t^{ipt}$ (Update biased variance estimate)

  $\hat{v}_t = \sigma_{U,t}^2 \leftarrow v_t + \beta_2^t(v_t - \frac{D}{2})$ (Compute robust bias-corrected variance estimate)

**end for**

**return** $\mu_{U,t}, \sigma_{U,t}^2$ at timesteps $t = 1, \cdots, T_0 + T_{Ada}$ (Estimating the first two moments)

Returning to the objective of convergence diagnostic-based outlier elimination, given that we are now capable of accurately estimating the minimum value $min\{U|c\}$ and variance $\sigma_U^2$ of the sample potential energy, and referencing the potential energy distribution of a Gaussian distribution, we set the threshold criterion for convergence diagnostic as follows:

$$U > min\{U|c\} + \sigma_U^2 + \lambda\sigma_U \qquad (47)$$

Here, the term $min\{U|c\} + \sigma_U^2$ provides a robust estimate of the mean of the sample potential energy $\mu_U$, while the term $\lambda\sigma_U$ aligns with the standard score method. Two values for the threshold parameter $\lambda$, namely $\lambda = 6$ and $\lambda = 50$, are recommended for different stages of the sampling process. The outliers from non-converged samples satisfying Eq. (47) will not participate in the adaptive estimation, i.e., each batch mean $\frac{1}{K}\sum_{k=1}^{K} \cdot$ in Algorithms 2 and 3 will be replaced by the mean calculated from the remaining acceptable samples. In the applications of Bayesian inference of structural dynamic models, we have observed that adopting a very lenient threshold with $\lambda = 50$ facilitates accurate direction estimation, whereas a slightly stricter threshold with $\lambda = 6$ aids in precise scale estimation.

An additional advantage of performing convergence diagnostic is enabling convergence acceleration and divergence mitigation in Markov chains. Prior to the application of adaptive PC estimates for normalizing the sampling process, the convergence of Markov chains tends to be relatively slow. Even after the normalization are applied, since the APM-SGHMC algorithm discards the parameter category $Cate_i$ from the network input, it cannot learn more targeted convergence strategies tailored to the variations in potential energy across different directions. Consequently, the convergence strategies trained under such circumstances will be relatively conservative. In such scenarios, it is appropriate to appropriately increase the step size $\eta$ to expedite convergence.

Therefore, for non-converged Markov chains detected by Eq. (47), we propose introducing a step-size relaxation coefficient matrix $\pmb{\Lambda}_\eta = \text{diag}(\pmb{\lambda}_\eta)$, where $\pmb{\lambda}_\eta \in \mathbb{R}^D$. This allows us to rewrite the discretized update rules in Eqs. (14)-(15) as follows:

$$\pmb{p}_{t+1} = \left(1 - \eta \pmb{\Lambda}_\eta \pmb{C}(\pmb{z}_t)\right) \pmb{p}_t - \eta \pmb{\Lambda}_\eta \pmb{G}(\pmb{z}_t) \pmb{\nabla}_{\pmb{\theta}_t} U(\pmb{\theta}_t)$$

$$+ \eta \pmb{\Lambda}_\eta \left[\left(\pmb{\nabla}_{\pmb{\theta}_t} \cdot \pmb{G}^T(\pmb{z}_t)\right)^T + \left(\pmb{\nabla}_{\pmb{p}_t} \cdot \pmb{C}^T(\pmb{z}_t)\right)^T\right] + \sqrt{2\eta} \pmb{\Lambda}_\eta^{1/2} \pmb{\epsilon}_t \qquad (48)$$

$$\pmb{\theta}_{t+1} = \pmb{\theta}_t + \eta \pmb{\Lambda}_\eta \pmb{G}(\hat{\pmb{z}}_t) \pmb{p}_{t+1} - \eta \pmb{\Lambda}_\eta \left(\pmb{\nabla}_{\pmb{p}_{t+1}} \cdot \pmb{G}^T(\hat{\pmb{z}}_t)\right)^T \qquad (49)$$

For non-converged Markov chains, since their samples are unlikely to conform to the stationary distribution in the short term, our primary concern is to ensure they do not exhibit a tendency to diverge. Therefore, we initially set the relaxation coefficient uniformly to a relatively large value, for example, $\pmb{\Lambda}_\eta = 3\mathbf{I}_D$, and then monitor their divergence tendency. To this end, we propose two empirical criteria for divergence

detection. For the entire Markov chain, the criterion for divergence tendency is defined as:

$$U(\boldsymbol{\theta}_t) - U(\boldsymbol{\theta}_{t-1}) > 10\sqrt{D/2} \text{ and } U(\boldsymbol{\theta}_t) > U(\boldsymbol{\theta}_{t-2}) \tag{50}$$

Whenever a non-converged Markov chain satisfies Eq. (50), we reset its momentum to zero, $\boldsymbol{p}_{t+1} \leftarrow \boldsymbol{0}$, and simultaneously reduce the relaxation coefficient element-wise as follows:

$$\lambda_\eta \leftarrow \hat{\lambda}_{min} \left( \frac{\lambda_\eta}{\hat{\lambda}_{min}} \right)^\gamma \tag{51}$$

where $\hat{\lambda}_{min} = 1/3$ and $\gamma = 0.8$ are recommended; $\hat{\lambda}_{min}$ is a sufficiently small relaxation coefficient that plays a positive role in the convergence of the chain and $\gamma \in (0,1)$ is the decay rate of the reduction. It can be observed that after multiple reductions according to Eq. (51), the relaxation coefficient will tend towards $\hat{\lambda}_{min}$. For the $i$-th state component $\theta_i$ within the Markov chain, the criterion for divergence tendency is given by:

$$|p_{t,i}| > 5 \text{ and } p_{t,i} \partial_{\theta_i} U(\boldsymbol{\theta}_{t-1}) > 0 \tag{52}$$

This implies that the sample is moving with a significant velocity in a direction where the probability decreases, typically indicating that it has reached a local optimum in that component direction. If a chain satisfies Eq. (52) for the $i$-th state component $\theta_i$, we reset its momentum in that direction to zero, $p_{t+1,i} \leftarrow 0$. Moreover, for every three cumulative instances where this condition is met, we reduce the relaxation coefficient in the corresponding direction as follows:

$$\lambda_{\eta,i} \leftarrow \hat{\lambda}_{min} \left( \frac{\lambda_{\eta,i}}{\hat{\lambda}_{min}} \right)^\gamma \tag{53}$$

where $\hat{\lambda}_{min} = 1/3$ and $\gamma = 0.8$ are recommended, as in Eq. (51).

For instance, when utilizing the APM-SGHMC algorithm in subsequent examples,

we adopted the following settings: During the initial burn-in phase spanning the first $T_b = 3000$ steps for Section 5.1 or $T_b = 2000$ for Section 5.2, we commenced with a convergence diagnostic threshold of $\lambda = 50$. For all chains, we reduced the energy input to the particle system during the simulation process by setting $\epsilon_t \leftarrow 0.5\epsilon_t$. The step-size relaxation coefficient matrix for all chains was initialized as $\Lambda_\eta = 10\mathbf{I}_D$, to be activated for non-converged chains. At $T_0 = 200$ steps, we initiated the adaptive estimation algorithm for statistical purposes, but its results were not yet applied to the sampling process. At $T_r = 300$ steps, we reset the step-size relaxation coefficient matrix for all chains to $\Lambda_\eta = 3\mathbf{I}_D$ and commenced sampling under rotational transformations using the adaptive PC estimation results. However, the scale adaptive estimates were scaled down as $\{\sigma_{d,t}\}_{d=1}^D \leftarrow \{0.5\sigma_{d,t}\}_{d=1}^D$, which rapidly reduced the sample potential energy below the threshold, thereby roughly determining the PCDs. At $T_f = 500$ steps, we once again reset the step-size relaxation coefficient matrix for all chains to $\Lambda_\eta = 3\mathbf{I}_D$ and simultaneously changed the convergence diagnostic threshold to $\lambda = 6$. This ensured that the sample potential energy in the remaining non-converged chains rapidly descended to near the minimum value, with the scale adaptive estimates stabilizing at a position slightly below the true values. At $T_s = 800$ steps, we resumed normal sampling, meaning we no longer reduced the energy input or scaled down the scale adaptive estimates. Concurrently, we reverted the convergence diagnostic threshold to $\lambda = 50$, implying that essentially no non-converged samples would be detected, thereby rendering the adaptive estimation results more precise. At $T_0 + T_{Ada} = T_b - 200$ steps, we terminated the adaptive estimation, and the estimated values were fixed as sampling parameters, no longer subject to any change.

## 4. APM-SGHMC ALGORITHM

According to the previous description, the APM-SGHMC algorithm is summarized in Algorithm 4. Moreover, Fig. 2 further simplifies it into a conceptual flowchart, illustrating the correlation between the main algorithmic components.

---

**Algorithm 4:** APM-SGHMC algorithm

---

**Require:** $U_w(\cdot)$ and $\nabla_w U_w(\cdot)$: Functions for computing potential energy and its gradient.
**Require:** $T$, $T_b$, $T_0$, $T_r$, $T_f$, $T_s$, $T_{Ada}$ and $N$: Key time-steps controlling the algorithm.
**Require:** $\eta$ and $K$: Step size and batch size of samples

$\{z_{0,k}\}_{k=1}^{K} = \{(\boldsymbol{\theta}_{0,k}, \boldsymbol{p}_{0,k})\}_{k=1}^{K} \leftarrow \{(\mathbf{0}_D, \mathbf{0}_D)\}_{k=1}^{K}$ (Initialize augmented state vectors)

$\{\mu_{U,0}, \sigma_{U,0}^2\} \leftarrow \{0, \frac{D}{2}\}$ (Initialize moment estimate of the potential energy)

$\{\boldsymbol{P}_0, \hat{\boldsymbol{v}}_0 = (\sigma_{1,0}^2, \cdots, \sigma_{D,0}^2)^T\} \leftarrow \{\boldsymbol{I}_D, \boldsymbol{v}_0^*\}$ (Initialize PC estimate of the parameter vector)

$\left\{\left\{\lambda_{\eta,i}^{(0,k)}\right\}_{i=1}^{D}\right\}_{k=1}^{K} \leftarrow \{\{10\}_{i=1}^{D}\}_{k=1}^{K}$ (Initialize relaxation coefficients for non-converged chains)

$t \leftarrow 0$ (Initialize timestep)
$t_0 \leftarrow 0$ (Initialize timestep for back-propagation check)
**for** $t' = 1, \cdots, N$ **do**

    Compute $\{U(\boldsymbol{\theta}_{t,k})\}_{k=1}^{K} \leftarrow \{U_w(\boldsymbol{P}_t \boldsymbol{\theta}_{t,k})\}_{k=1}^{K}$

    Detect *outliers* (non-converged samples) as in Eq. (47)
    **if** $t = T_r$ or $t = T_f$ **do**

        $\left\{\left\{\lambda_{\eta,i}^{(0,k)}\right\}_{i=1}^{D}\right\}_{k=1}^{K} \leftarrow \{\{3\}_{i=1}^{D}\}_{k=1}^{K}$

    **for** $k = 1, \cdots, K$ **do**
        **if** $\boldsymbol{\theta}_{t,k}$ is *outlier* **do**

            Update $\boldsymbol{\lambda}_\eta^{(t+1,k)} = \left(\lambda_{\eta,1}^{(t+1,k)}, \cdots, \lambda_{\eta,D}^{(t+1,k)}\right)^T$ as in Eqs. (50)-(53)

        $\boldsymbol{\Lambda}_\eta^{(t+1,k)} \leftarrow diag(\boldsymbol{\lambda}_\eta^{(t+1,k)})$
        **else do**

            $\left\{\left\{\lambda_{\eta,i}^{(t+1,k)}\right\}_{i=1}^{D}\right\}_{k=1}^{K} \leftarrow \left\{\left\{\lambda_{\eta,i}^{(t,k)}\right\}_{i=1}^{D}\right\}_{k=1}^{K}$

        $\boldsymbol{\Lambda}_\eta^{(t+1,k)} \leftarrow \boldsymbol{I}_D$
    **end for**
    **if** $t < T_0 + T_{Ada}$ **do**

        Compute $\mu_{U,t+1}, \sigma_{U,t+1}^2$ as in Algorithm 3 with input $\{U(\boldsymbol{\theta}_{t,k})\}_{k=1}^{K}$ ignoring *outliers*

    **else do**
        $\{\mu_{U,t+1}, \sigma_{U,t+1}^2\} \leftarrow \{\mu_{U,t}, \sigma_{U,t}^2\}$
    **if** $T_0 \leq t < T_0 + T_{Ada}$ **do** (Adaptive PC estimation)

**if** $t < T_r$ **do**

    Compute $\{\boldsymbol{\theta}_{t,k}^{\text{waste}}\}_{k=1}^{K}, \boldsymbol{P}_{t+1}, \widehat{\boldsymbol{v}}_{t+1}^{\text{waste}}$ as in Algorithm **2** with input $\{\boldsymbol{P}_t^T \boldsymbol{\theta}_{t,k}\}_{k=1}^{K}$ ignoring *outliers*

$$\widehat{\boldsymbol{v}}_{t+1} = (\sigma_{1,t+1}^2, \cdots, \sigma_{D,t+1}^2)^T \leftarrow \widehat{\boldsymbol{v}}_t$$

**else do**

    **if** $t = T_r$ **do**

        $\{\boldsymbol{\theta}_{t,k}, \boldsymbol{p}_{t,k}\}_{k=1}^{K} \leftarrow \{\boldsymbol{P}_t^T \boldsymbol{\theta}_{t,k}, \boldsymbol{P}_t^T \boldsymbol{p}_{t,k}\}_{k=1}^{K}$ (Start adaptive rotation of state space)

        Compute $\{\boldsymbol{\theta}_{t,k}\}_{k=1}^{K}, \boldsymbol{P}_{t+1}, \widehat{\boldsymbol{v}}_{t+1} = (\sigma_{1,t+1}^2, \cdots, \sigma_{D,t+1}^2)^T$ as in Algorithm **2** with input $\{\boldsymbol{\theta}_{t,k}\}_{k=1}^{K}$ ignoring *outliers*

        $\{\boldsymbol{p}_{t,k}\}_{k=1}^{K} \leftarrow \{\boldsymbol{P}_{t+1}^T \boldsymbol{P}_t \boldsymbol{p}_{t,k}\}_{k=1}^{K}$ (Adaptive rotation of momentum samples)

    **else do**

        $\{\boldsymbol{P}_{t+1}, \widehat{\boldsymbol{v}}_{t+1} = (\sigma_{1,t+1}^2, \cdots, \sigma_{D,t+1}^2)^T\} \leftarrow \{\boldsymbol{P}_t, \widehat{\boldsymbol{v}}_t\}$

**if** $t < T_s$ **do**

    $\{\sigma_{d,t+1}\}_{d=1}^{D} \leftarrow \{0.5\sigma_{d,t+1}\}_{d=1}^{D}$

Compute $\{\nabla_{\boldsymbol{\theta}} U(\boldsymbol{\theta}_{t,k})\}_{k=1}^{K} \leftarrow \{\boldsymbol{P}_{t+1}^T \nabla_{\boldsymbol{w}} U_{\boldsymbol{w}}(\boldsymbol{P}_{t+1} \boldsymbol{\theta}_{t,k})\}_{k=1}^{K}$

$\{\widehat{U}(\boldsymbol{\theta}_{t,k}), \nabla_{\boldsymbol{\theta}} \widehat{U}(\boldsymbol{\theta}_{t,k})\}_{k=1}^{K} \leftarrow \left\{\frac{U(\boldsymbol{\theta}_{t,k}) - \mu_{U,t+1}}{\sqrt{2D}\sigma_{U,t+1}}, \frac{\nabla_{\boldsymbol{\theta}} U(\boldsymbol{\theta}_{t,k})}{\sqrt{2D}\sigma_{U,t+1}}\right\}_{k=1}^{K}$

Compute $\nabla_{\boldsymbol{\theta}} \widehat{U}^*(\boldsymbol{\theta}_{t,k})$ by $\{\partial_{\theta_i} \widehat{U}^*(\boldsymbol{\theta}_{t,k})\}_{i=1}^{D} \leftarrow \{\sigma_{i,t+1} \cdot \partial_{\theta_i} \widehat{U}(\boldsymbol{\theta}_{t,k})\}_{i=1}^{D}$

$\{\boldsymbol{G}(\boldsymbol{z}_{t,k}), \boldsymbol{C}(\boldsymbol{z}_{t,k})\}_{k=1}^{K} \leftarrow \{\boldsymbol{G}_{\phi_Q}(\widehat{U}(\boldsymbol{\theta}_{t,k}), \boldsymbol{p}_{t,k}), \boldsymbol{C}_{\phi_D}(\widehat{U}(\boldsymbol{\theta}_{t,k}), \boldsymbol{p}_{t,k}, \nabla_{\boldsymbol{\theta}} \widehat{U}^*(\boldsymbol{\theta}_{t,k}))\}_{k=1}^{K}$ (Invoke NNs)

Generate $\boldsymbol{\epsilon}_{t,k} \sim \mathcal{N}(\boldsymbol{0}_D, \boldsymbol{C}(\boldsymbol{z}_{t,k}))$ for $k = 1, \cdots, K$

**if** $t < T_s$ **do**

    $\{\boldsymbol{\epsilon}_{t,k}\}_{k=1}^{K} \leftarrow \{0.5\boldsymbol{\epsilon}_{t,k}\}_{k=1}^{K}$

$\{\boldsymbol{p}_{t+1,k}\}_{k=1}^{K} \leftarrow \Big\{\big(1 - \eta \Lambda_\eta^{(t+1,k)} \boldsymbol{C}(\boldsymbol{z}_{t,k})\big) \boldsymbol{p}_{t,k} - \eta \Lambda_\eta^{(t+1,k)} \boldsymbol{G}(\boldsymbol{z}_{t,k}) \nabla_{\boldsymbol{\theta}} U(\boldsymbol{\theta}_{t,k}) + \eta \Lambda_\eta^{(t+1,k)} \big[(\nabla_{\boldsymbol{\theta}} \cdot \boldsymbol{G}^T(\boldsymbol{z}_{t,k}))^T + (\nabla_{\boldsymbol{p}} \cdot \boldsymbol{C}^T(\boldsymbol{z}_{t,k}))^T\big] + \sqrt{2\eta} \Lambda_\eta^{(t+1,k)1/2} \boldsymbol{\epsilon}_{t,k}\Big\}_{k=1}^{K}$

$\{\widehat{\boldsymbol{z}}_{t,k}\}_{k=1}^{K} \leftarrow \{(\boldsymbol{\theta}_{t,k}, \boldsymbol{p}_{t+1,k})\}_{k=1}^{K}$ (Denote)

$\{\boldsymbol{G}(\widehat{\boldsymbol{z}}_{t,k})\}_{k=1}^{K} \leftarrow \{\boldsymbol{G}_{\phi_Q}(\widehat{U}(\boldsymbol{\theta}_{t,k}), \boldsymbol{p}_{t+1,k})\}_{k=1}^{K}$ (Invoke NNs)

$$\{\boldsymbol{\theta}_{t+1,k}\}_{k=1}^{K} \leftarrow \left\{\boldsymbol{\theta}_{t,k} + \eta \boldsymbol{\Lambda}_\eta^{(t+1,k)} \boldsymbol{G}(\hat{\boldsymbol{z}}_{t,k})\boldsymbol{p}_{t+1,k} - \eta \boldsymbol{\Lambda}_\eta^{(t+1,k)} \left(\boldsymbol{\nabla}_{\boldsymbol{p}} \cdot \boldsymbol{G}^{\mathrm{T}}(\hat{\boldsymbol{z}}_{t,k})\right)^{\mathrm{T}}\right\}_{k=1}^{K}$$ (Obtain new samples)

$$\{\boldsymbol{z}_{t+1,k}\}_{k=1}^{K} \leftarrow \{(\boldsymbol{\theta}_{t+1,k}, \boldsymbol{p}_{t+1,k})\}_{k=1}^{K}$$ (Denote)

$t \leftarrow t + 1$

**if** training **do**

    Recreate a copy of $\{\boldsymbol{z}_{t,k}\}_{k=1}^{K}$ to stop the gradient flow;

    **if** $t = t_0 + T$ **do**

        Update the NNs by back-propagation through the loss function Eq. (16) with samples $\{\{\boldsymbol{\theta}_{s,k}\}_{s=t_0+1}^{t_0+T}\}_{k=1}^{K}$.

    $t_0 \leftarrow t$

**end for**

**return** $\boldsymbol{P}_N$ (Estimated PCD vectors)

**return** $\{\boldsymbol{\theta}_{t,k}\}_{k=1}^{K}$ at timesteps $t = T_b + 1, T_b + 2, \cdots, N$ (Generated state samples)

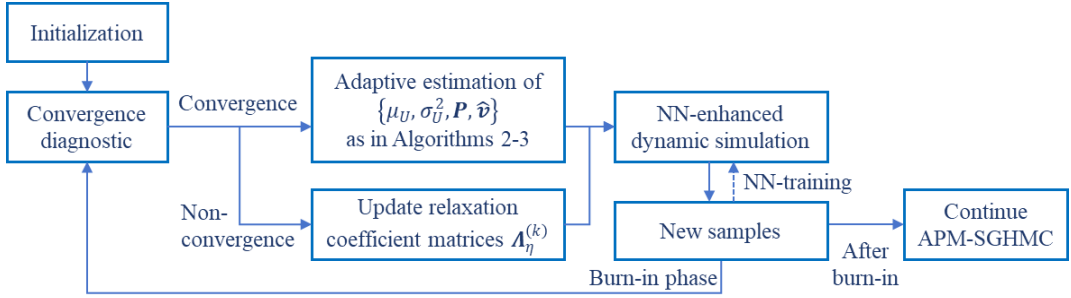

**Figure 2.** Conceptual flowchart of APM-SGHMC algorithm.

## 5. ILLUSTRATIVE EXAMPLES

### *5.1 Bayesian system identification of building structural model*

In this section, for comparison with AM-SGHMC, we first consider an example from [31], namely, *N*-story buildings subjected to earthquake excitation. This example refers to the benchmark structure [37-40] proposed by IASC-ASCE (International Association for Structural Control-American Society of Civil Engineers) Task Group on SHM. The 4-story 2-bay by 2-bay steel braced-frame benchmark test structure is shown in Fig. 3. The $N$-story buildings in this example adopt the same floor structural layout. The planar motion of 3 degrees of freedom (DOF) for each floor is considered, so that $3N$-DOF models are utilized for Bayesian inference and the Young's modulus parameters $E_i^{(j)}$ for 5 regions, $j = 1,\cdots,5$, as detailed in [31], of each story, $i = 1,\cdots,N$, of the models are estimated. Altogether, we need to estimate $D = 5N + 1$ model parameters with error parameter $\sigma$ included. The likelihood function $p(\mathcal{D}|\mathbf{w})$ follows the same form as in Eq. (2), with $N_o = 8$, $N_T = 300$.

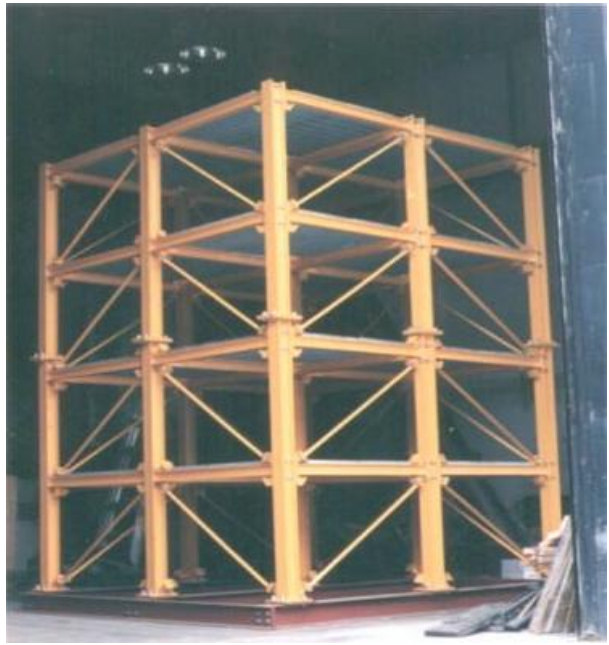

**Figure 3.** The 4-story IASC-ASCE benchmark test structure [37].

Based on the nominal value $E_0 = 2 \times 10^{11}\text{Pa}$ (not equal to the exact value), and a roughly estimated error magnitude $\sigma_0 = 0.8\text{ms}^{-2}$, the uncertain parameters $w_k, k = 1,\cdots,5N + 1$, are defined in dimensionless form: $w_{i\times5+j-5} = E_i^{(j)}/E_0$ for $i = 1,\cdots,N; j = 1,\cdots,5$; and $w_{5N+1} = \sigma/\sigma_0$. We set the prior PDF $p(\mathbf{w})$ as truncated independent distributions. The dimensionless modulus of each component determined by parameters $w_k = E_i^{(j)}/E_0$, $k = 1,\cdots,5N$ follow Gaussian distributions with means of 1 and coefficients of variation (c.o.v.) of 10%, truncated by interval

[0.1,2.0], and $w_{5N+1} = \sigma/\sigma_0$ follows a lognormal distribution with median 1 and a logarithmic standard deviation of $s_0 = 0.3$ (the c.o.v. is about 30%), also truncated by interval [0.1,2.0]. The prior boundaries are transformed as in [31]. Then the potential energy $U_w(w)$ can be calculated as stated following Eq. (20).

For APM-SGHMC, the architecture of two embedded NNs $f_{\phi_Q}(\cdot)$ and $f_{\phi_D}(\cdot)$ and training setup are the same as [31] except that the NN input $Cate_i$ is discarded, as detailed in Appendix D. The experimental arrangement for this example is shown in Fig. 4. The sampler is firstly trained by sampling on Dataset 1 (4-story with typical noise), and then tested on Datasets 2-4 (2-story, 4-story and 6-story with large noise as generalization tasks). These four sets of noisy accelerometer data are all collected from the base, the first floor, and the roof, along the horizontal direction of each outer wall, with a sample interval of 0.01s and a duration of 3 s, as detailed in [31].

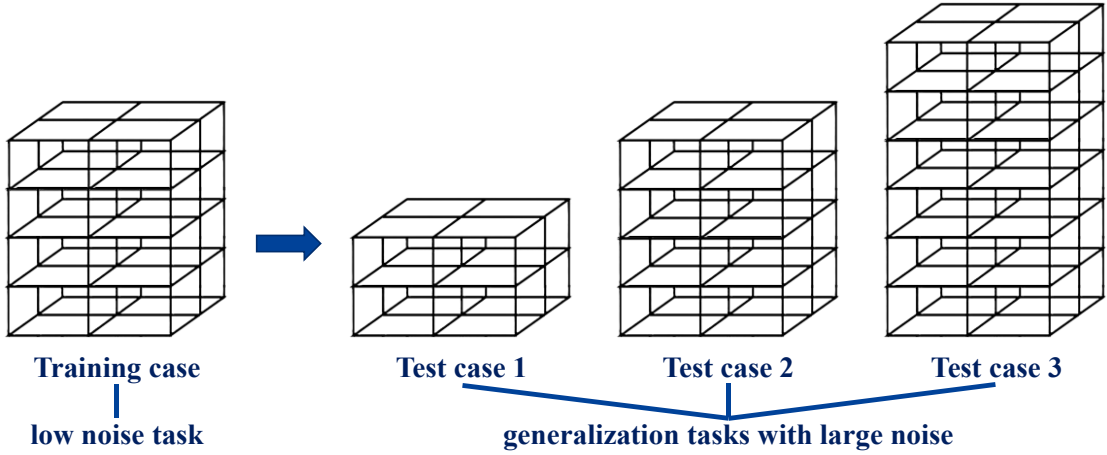

**Figure 4.** Experimental arrangement for multi-story braced-frame building example.

During each test of APM-SGHMC, $K = 32$ parallel chains for $N = 9000$ steps are simulated with step-size $\eta = \sqrt{0.001}$. The adaptive estimates are updated only during steps 200 to 1800 with $\left(t_1^b, \hat{t}_{1,min}, \hat{t}_{1,max}\right)_\theta = (100,100,1000)$, $\left(t_2^b, \hat{t}_{2,min}, \hat{t}_{2,max}\right)_\theta = (200,200,1000)$, $\left(t_3^b, \hat{t}_{3,min}, \hat{t}_{3,max}\right)_\theta = (600,600,1000)$ and $(\beta_1, \beta_2)_U = (0.98, 0.99)$ for parameter samples $\boldsymbol{\theta}$ and potential energy $U(\boldsymbol{\theta})$. The samples (after burn-in $T_b = 3000$) for three cases by APM-SGHMC for last two PCs of parameters are shown in Fig. 5 and compared with those by AM-SGHMC and HMC method from [31]. For each test of AM-SGHMC, $K = 32$ parallel chains for $N = 9000$ steps are simulated, where the first $T_b = 3000$ steps are burn-in. For each test of HMC, the initial point is determined by 4000 steps of an efficient SPSA (simultaneous perturbation stochastic approximation) optimization algorithm, then $K = 32$ parallel chains for $N = 9000$ steps are simulated, where the first $T_b = 500$ steps are burn-in. The colors in the figures are the two-dimensional marginal PDFs of the posterior PDFs estimated from the samples. As shown in the figure, the posterior sample distributions obtained by APM-SGHMC and HMC method exhibit similar patterns, indicating that APM-SGHMC generated reliable samples across these two directions. While AM-SGHMC covers a larger area, indicating that it generated more samples with slightly lower posterior PDFs.

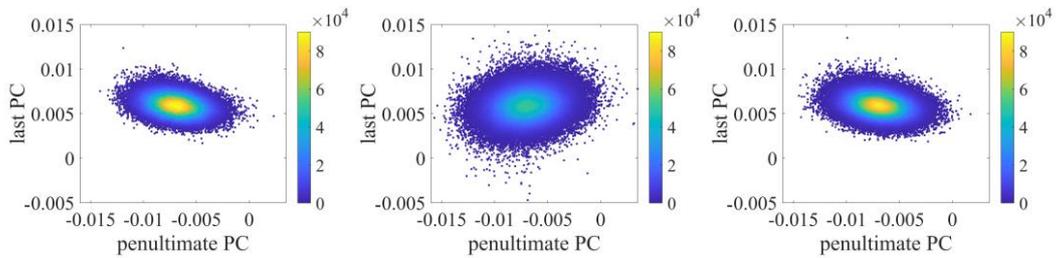

(a) Samples of Dataset 2 (2-story).

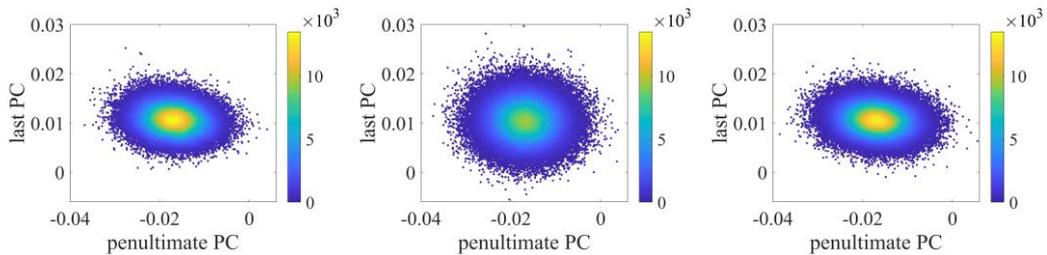

(b) Samples of Dataset 3 (4-story).

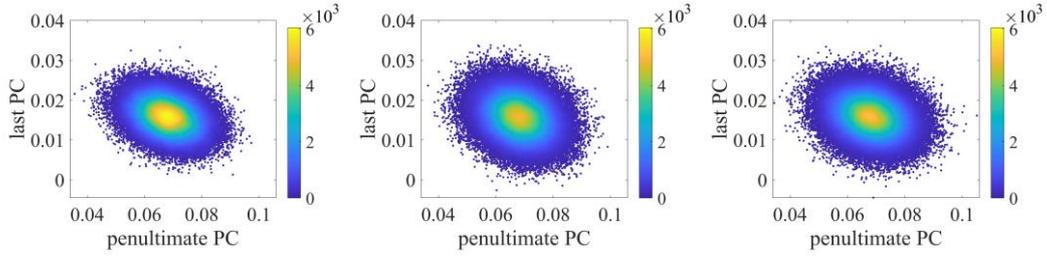

(c) Samples of Dataset 4 (6-story).

**Figure 5.** Sample plots for last two PCs of parameters by HMC (left column), AM-SGHMC (middle column) and APM-SGHMC (right column) for three test cases.

The samples for three cases by APM-SGHMC for first two PCs of parameters are also shown in Fig. 6 and compared with those by AM-SGHMC and HMC method. The blue polylines in each figure are 160 Markov chains (5 repeated experiments, each comprising 32 parallel chains), with one of them highlighted in yellow. As can be seen from the figure, these two directions are not fully explored by a single chain of HMC method. AM-SGHMC accelerates the exploration of each parameter, thereby accelerating the exploration of first two PCs, but it is still insufficient. APM-SGHMC directly accelerates the exploration of each PC through adaptive rotation transformation so that the first two PCs can be explored as fully as the last two PCs.

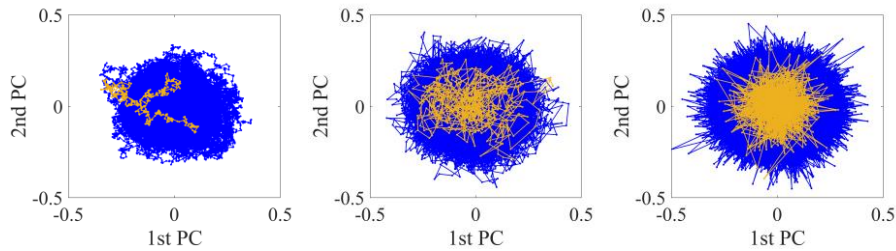

(a) Samples of Dataset 2 (2-story).

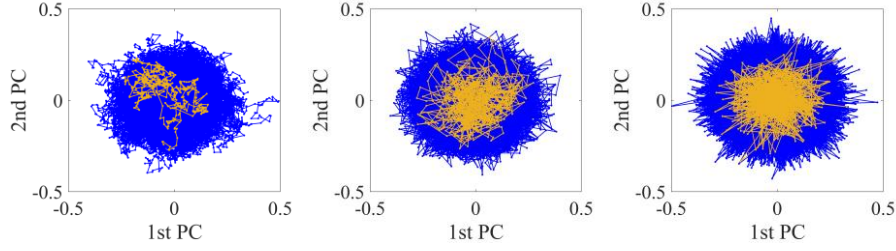

(b) Samples of Dataset 3 (4-story).

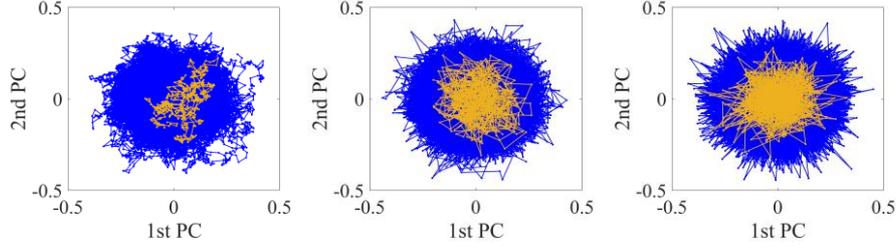

(c) Samples of Dataset 4 (6-story).

**Figure 6.** Sample plots for first two PCs of parameters by HMC (left column), AM-SGHMC (middle column) and APM-SGHMC (right column) for three test cases (Each subfigure displays 160 Markov chains with one highlighted).

To further compare the sampling results of the three methods, we use the negative ELBO, which is an approximation of the KLD between the estimated PDF of the samples by kernel density estimation and the posterior PDF, but with an unknown constant added. The negative ELBO, denoted as "KLD + c", of each Markov chain generated by the three methods on Datasets 2-4 are evaluated and their mean values are recorded in Tables 1-3, respectively. It can be seen that the negative ELBO of APM-SGHMC is always the smallest, which means that the distribution of the samples generated by APM-SGHMC fits the posterior PDF best. Based on these KLD results, the phenomenon depicted in Figs. 5-6 can be further explained as follows: Under the constraint of unknown correlations, the HMC method sacrifices exploration efficiency to ensure sample accuracy, whereas AM-SGHMC trades off some sample accuracy for higher exploration efficiency.

In order to compare the efficiency of the different samplers, Effective Sample Size (ESS) is employed, which is often used in the evaluation of samplers. ESS for a sequence of correlated samples can be viewed as the number of independent samples generated by the same PDF, where the two sets of samples can achieve the same accuracy in estimating the mean of the PDF. Our implementation of ESS estimation follows [26, 31] and is briefly presented here.

For a set of correlated samples $\boldsymbol{\Theta}_T = \{\boldsymbol{\theta}_\tau\}_{\tau=1}^T$ generated by a Markov chain, its ESS is estimated by:

$$ESS(\boldsymbol{\Theta}_T) = \frac{T}{1 + 2\sum_{s=1}^{\left[\frac{T}{3}\right]-1}\left(1 - \frac{s}{T}\right)\frac{\rho_s}{\rho_0}} \tag{54}$$

where $\frac{\rho_s}{\rho_0}$ is the autocorrelation of $\boldsymbol{\Theta}_T$ at lag $s$ estimated by:

$$\rho_s = \frac{1}{T-s}\sum_{\tau=s+1}^T (\boldsymbol{\theta}_\tau - \widehat{\boldsymbol{\mu}})^T(\boldsymbol{\theta}_{\tau-s} - \widehat{\boldsymbol{\mu}}) \tag{55}$$

where $\widehat{\boldsymbol{\mu}}$ is estimated by:

$$\widehat{\boldsymbol{\mu}} = \frac{1}{T}\sum_{\tau=1}^T \boldsymbol{\theta}_\tau \tag{56}$$

The sum in Eq. (54) is truncated whenever $s = 1000$ or $s$ is even number and $\rho_{s-1} + \rho_s < 0$.

The ESSs for the three methods on each Dataset are evaluated. With each corresponding time consumption recorded, we can define the sampling efficiency of the samplers as ESS per hour (ESS/h). The sampling efficiency of the three methods on Datasets 1-3 are also tabulated in Tables 1-3, respectively. As for the time consumption, all experiments in Section 5.1 and Section 5.2 are implemented on a computer equipped with an Intel i7-9700 K CPU (3.60 GHz, 8 cores), 16.0 GB RAM and an NVIDIA RTX

2080Ti GPU. In this section, both the MCMC algorithms and structural model computations were performed entirely on the GPU using PyTorch [41] in Python.

As can be seen from the tables, the efficiency of APM-SGHMC is 265, 174 and 171 times that of HMC, with 39, 21 and 19 times that of AM-SGHMC, in 2-story, 4-story and 6-story tasks, respectively. These results show that APM-SGHMC has the generalization ability and the almost identical ESS values of APM-SGHMC in these cases confirm the significance of eliminating the linear correlations between parameters through the use of rotation-invariant PCs.

**Table 1.** Sampling performance comparison of Dataset 2 (2-story).

| Methods | KLD + c | ESS | Time (h) | ESS/h |
| --- | --- | --- | --- | --- |
| HMC | 2685.80 | 7.96 | 3.20 | 2.49 |
| AM-SGHMC | 2684.13 | 31.32 | 1.89 | 16.59 |
| APM-SGHMC | **2682.84** | **1162.47** | 1.76 | **660.49** |

**Table 2.** Sampling performance comparison of Dataset 3 (4-story).

| Methods | KLD + c | ESS | Time (h) | ESS/h |
| --- | --- | --- | --- | --- |
| HMC | 2910.55 | 13.37 | 5.69 | 2.35 |
| AM-SGHMC | 2907.41 | 60.33 | 3.18 | 18.99 |
| APM-SGHMC | **2906.46** | **1165.49** | 2.84 | **410.38** |

**Table 3.** Sampling performance comparison of Dataset 4 (6-story).

| Methods | KLD + c | ESS | Time (h) | ESS/h |
| --- | --- | --- | --- | --- |
| HMC | 3001.90 | 14.24 | 8.29 | 1.71 |
| AM-SGHMC | 2994.92 | 69.89 | 4.59 | 15.23 |
| APM-SGHMC | **2993.39** | **1166.91** | 3.99 | **292.45** |

## 5.2 Bayesian system identification of bridge structural model with building-trained sampler

To demonstrate the universal generalization capability, that is, the trained APM-SGHMC sampler can be directly applied to other distinct sampling tasks, this section employs a novel bridge finite element model (FEM) derived from a classical bridge benchmark in OpenSees [42] (Open System for Earthquake Engineering Simulation). Since the NN input in the AM-SGHMC method includes parameter categories $Cate_i$, which requires that the parameter categories of the new structural model must be a subset of those used during training, the AM-SGHMC sampler must be retrained for this section. However, the minute-scale computation time required for a single potential energy and gradient evaluation in this bridge model leads to an impractically long training duration (on the order of days) for the AM-SGHMC method, necessitating its exclusion from comparison in this section.

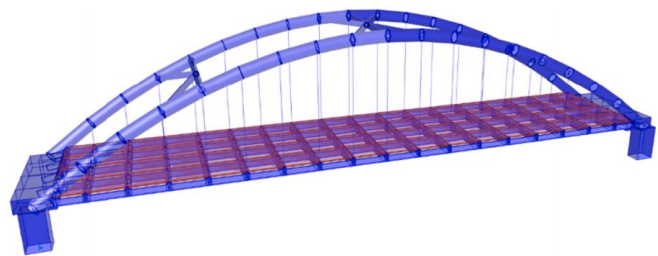

**Figure 7.** FEM of the bridge [43].

Fig. 7 shows the FEM of the bridge. The bridge system consists of 5 stiffening girders, 19 cross girders, 4 bridge piers, 2 arch ribs, 3 lateral bracings, and 34 hangers. The length and width of the bridge deck were 125 m and 24 m, respectively. The bridge was fixed on the ground with four piers. The height of the bridge pier was 7.425 m, and the cross-sectional area was $4m \times 3m$. The cross-sectional areas of the stiffening and cross girders were $1.2m \times 0.4m$ and $1.5m \times 1m$, respectively, and these components were constructed using reinforced concrete. The distance from the top chord to the bridge deck was 24.5 m. The arch ribs and lateral bracings were fabricated using circular hollow section steel tubes. The outer diameter and thickness of the arch rib were 2.5 and 0.5 m, respectively, while those of the lateral bracing were 1.2 and 0.4 m, respectively. The hangers were steel cables with a circular cross section of diameter 0.0713 m.

The FEM of the bridge was created in ETABS and converted to OpenSees for dynamic analysis [43]. Moreover, the OpenSees model is publicly accessible on http://www.dinochen.com/attachments/month_1305/020135702449.rar. To enable seamless integration with Python-based MCMC algorithms, the OpenSees model was further translated into OpenSeesPy [44] command syntax. The arch ribs, stiffening girders, cross girders, lateral bracings, and bridge piers were modeled using displacement-based beam-column elements according to their cross-section properties, while the slab was modeled using the shell element. The finite element mode of the bridge had 241 nodes and 439 elements. By employing the sensNodeAccel command in OpenSeesPy, we compute the structural response sensitivities, defined as the first-order partial derivatives of predicted structural responses with respect to model parameters, which are subsequently used to evaluate the potential energy gradient vector $\nabla_w U_w(w)$. In this section, since OpenSeesPy lacks GPU support, the sum of squared errors (SSE) between predicted and measured structural responses, along with their sensitivity derivatives, are computed in parallel across multiple CPU cores, while the remaining components of the MCMC algorithms continue to be performed on the GPU.

To better introduce the model parameterization, as illustrated in Fig. 8, the bridge model is divided into three longitudinal regions (I, II, III), with the front and rear elevations denoted as f and r, respectively. Structural components are abbreviated as follows: piers (P), arch ribs (A), hangers (H), stiffening and cross girders (G, including sG and cG) and lateral bracings (B). These abbreviations are used for Young's modulus parameter naming conventions in this section. Three model classes are considered. Based on the nominal values of Young's modulus for concrete $E_{c0} = 2.6 \times 10^{10}$ Pa and steel $E_{s0} = 2.06 \times 10^{11}$ Pa (not equal to the exact value), along with a roughly estimated error magnitude $\sigma_0 = 0.1$ m/s$^2$, the dimensionless uncertain parameters are selected and tabulated in Table 4.

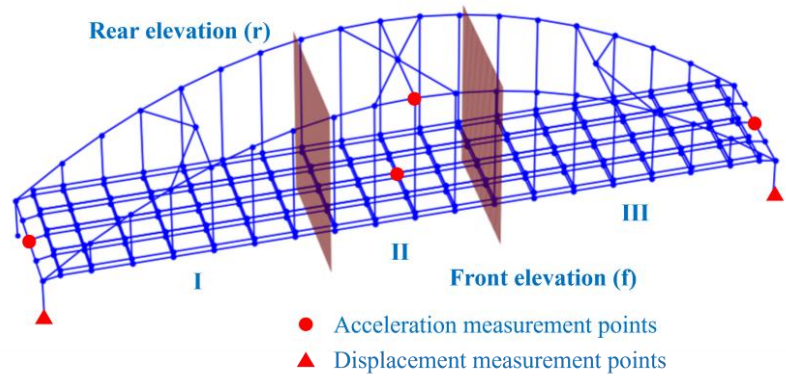

**Figure 8.** Structural region division and measurement point arrangement.

**Table 4.** Dimensionless uncertain parameters of three bridge model classes.

| Model class | Parameters |
| --- | --- |
| Model Class 1 (6 parameters) | $E^{(P)}/E_{c0}$, $E^{(A)}/E_{s0}$, $E^{(H)}/E_{s0}$, $E^{(G)}/E_{c0}$, $E^{(B)}/E_{s0}$ and $\sigma/\sigma_0$ |
| Model Class 2 (17 parameters) | $E_I^{(P)}/E_{c0}$, $E_{III}^{(P)}/E_{c0}$, $E_f^{(A)}/E_{s0}$, $E_r^{(A)}/E_{s0}$, $E_{I,f}^{(H)}/E_{s0}$, $E_{I,r}^{(H)}/E_{s0}$, $E_{II,f}^{(H)}/E_{s0}$, $E_{II,r}^{(H)}/E_{s0}$, $E_{III,f}^{(H)}/E_{s0}$, $E_{III,r}^{(H)}/E_{s0}$, $E_I^{(G)}/E_{c0}$, $E_{II}^{(G)}/E_{c0}$, $E_{III}^{(G)}/E_{c0}$, $E_I^{(B)}/E_{s0}$, $E_{II}^{(B)}/E_{s0}$, $E_{III}^{(B)}/E_{s0}$ and $\sigma/\sigma_0$ |
| Model Class 3 (29 parameters) | $E_{I,f}^{(P)}/E_{c0}$, $E_{I,r}^{(P)}/E_{c0}$, $E_{III,f}^{(P)}/E_{c0}$, $E_{III,r}^{(P)}/E_{c0}$, $E_{I,f}^{(A)}/E_{s0}$, $E_{I,r}^{(A)}/E_{s0}$, $E_{II,f}^{(A)}/E_{s0}$, $E_{II,r}^{(A)}/E_{s0}$, $E_{III,f}^{(A)}/E_{s0}$, $E_{III,r}^{(A)}/E_{s0}$, |

$E_{\mathrm{I,f}}^{(\mathrm{H})}/E_{s0}$, $E_{\mathrm{I,r}}^{(\mathrm{H})}/E_{s0}$, $E_{\mathrm{II,f}}^{(\mathrm{H})}/E_{s0}$, $E_{\mathrm{II,r}}^{(\mathrm{H})}/E_{s0}$, $E_{\mathrm{III,f}}^{(\mathrm{H})}/E_{s0}$, $E_{\mathrm{III,r}}^{(\mathrm{H})}/E_{s0}$,

$E_{\mathrm{I}}^{(\mathrm{sG})}/E_{c0}$, $E_{\mathrm{II}}^{(\mathrm{sG})}/E_{c0}$, $E_{\mathrm{III}}^{(\mathrm{sG})}/E_{c0}$,

$E_{\mathrm{I,f}}^{(\mathrm{cG})}/E_{c0}$, $E_{\mathrm{I,r}}^{(\mathrm{cG})}/E_{c0}$, $E_{\mathrm{II,f}}^{(\mathrm{cG})}/E_{c0}$, $E_{\mathrm{II,r}}^{(\mathrm{cG})}/E_{c0}$, $E_{\mathrm{III,f}}^{(\mathrm{cG})}/E_{c0}$, $E_{\mathrm{III,r}}^{(\mathrm{cG})}/E_{c0}$,

$E_{\mathrm{I}}^{(\mathrm{B})}/E_{s0}$, $E_{\mathrm{II}}^{(\mathrm{B})}/E_{s0}$, $E_{\mathrm{III}}^{(\mathrm{B})}/E_{s0}$

and $\sigma/\sigma_0$.

Same as the previous example, we set the prior PDF $p(\mathbf{w})$ as truncated independent distributions. Each dimensionless modulus follows a Gaussian distribution with a mean of 1 and a c.o.v. of 10%, and each $\sigma/\sigma_0$ follows a lognormal distribution with a median 1 and a logarithmic standard deviation $s_0 = 0.3$ (the c.o.v. is about 30%). All distributions are truncated to the interval $[0.1, 2.0]$. The likelihood function $p(\mathcal{D}|\mathbf{w})$ follows the same form as in Eq. (2), with $N_o = 6$, $N_T = 200$, a dataset $\mathcal{D}$ simulated using a model under Model Class 2 and three bridge model classes incorporated.

The dataset comprises noisy accelerometer data and ground motion displacement data, with a sample interval of 0.02 seconds and a duration of 4 seconds, as illustrated in Fig. 9. Displacement data comes from the bottom of the piers, measured longitudinally. Total acceleration data is collected from the top of the piers in the longitudinal direction, and also from the top of the front arch rib and the mid-span of the bridge deck, with measurements in both longitudinal (x) and vertical (z) directions. The specific locations of the measurement points are elaborated in Fig. 8. The experimental arrangement for this example is shown in Fig. 10. The APM-SGHMC sampler, having been trained in the previous example, is tested on this dataset across Model Class 1-3 (6-para, 17-para and 29-para) to evaluate generalization performance.

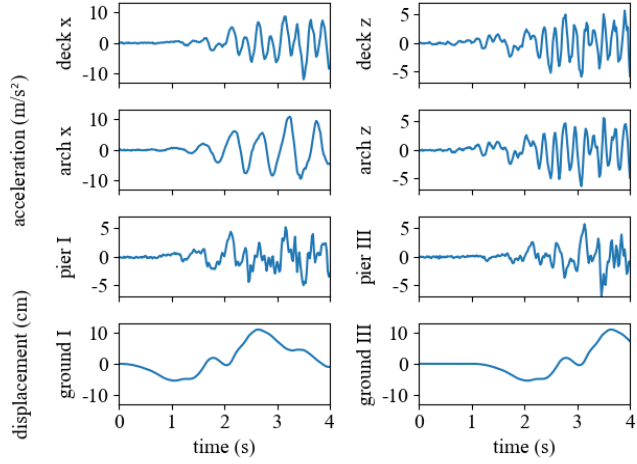

**Figure 9.** Dataset in bridge example.

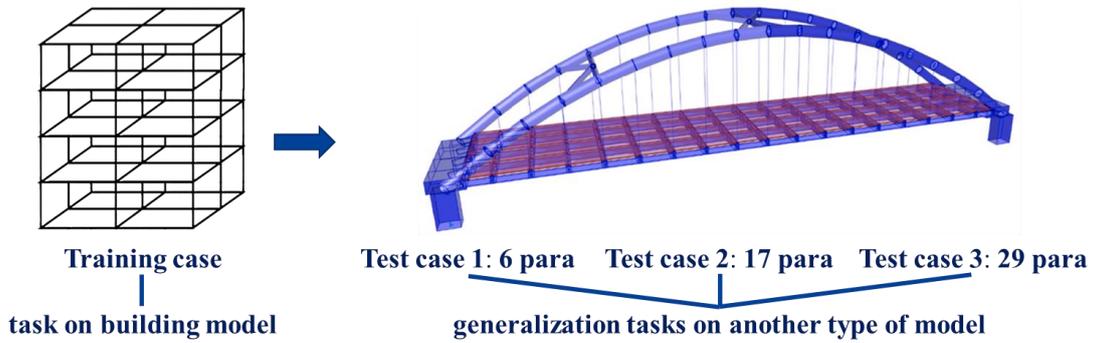

**Figure 10.** Experimental arrangement for bridge example.

During each test of APM-SGHMC, $K = 8$ parallel chains for $N = 4500$ steps are simulated with step-size $\eta = \sqrt{0.001}$. The configuration of adaptive estimates is the same as the previous example. The samples (after burn-in $T_b = 2000$) for three cases by APM-SGHMC for last two PCs of parameters are shown in Fig. 11 and compared with those by HMC method. For each test of HMC, the initial point is determined by 3000 steps of an efficient SPSA optimization algorithm, then $K = 8$ parallel chains for $N = 3600$ steps are simulated, where the first $T_b = 1100$ steps are burn-in. The colors in the figures are the two-dimensional marginal PDFs of the posterior PDFs estimated from the samples. As shown in the figure, similar to the previous example, the posterior sample distributions obtained by the two methods exhibit similar patterns, indicating that both methods generated reliable samples across these two directions. Regarding the central high-probability region obtained by HMC, the brighter coloration indicates nonlinear dependencies between PCs. Insufficient exploration of other PCs by HMC (shown later in Fig. 12) leads to its estimated marginal PDFs resembling concentrated conditional PDFs, rather than flatter true marginal PDFs formed by weighted averaging of continuously varying conditional PDFs across different values of those PCs [31].

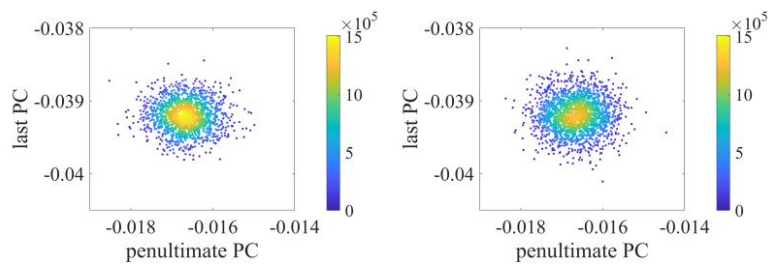

(a) Samples of Model Class 1 (6-para).

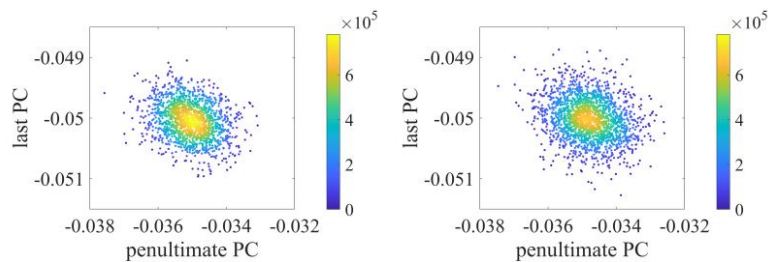

(b) Samples of Model Class 2 (17-para).

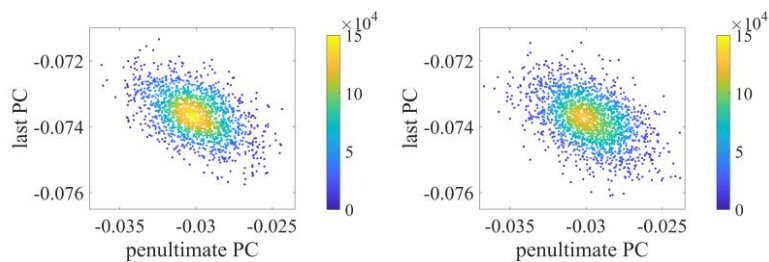

(c) Samples of Model Class 3 (29-para).

**Figure 11.** Sample plots for last two PCs of parameters by HMC (left column) and APM-SGHMC (right column) for three cases.

The samples for three cases by APM-SGHMC for first two PCs of parameters are also shown in Fig. 12 and compared with those by HMC method. Different colors of the samples in the figures correspond to different Markov chains. As can be seen from the figure, similar to the previous example, these two directions are not fully explored by HMC method. While APM-SGHMC directly accelerates the exploration of each PC through adaptive rotation transformation so that the first two PCs can be explored as fully as the last two PCs.

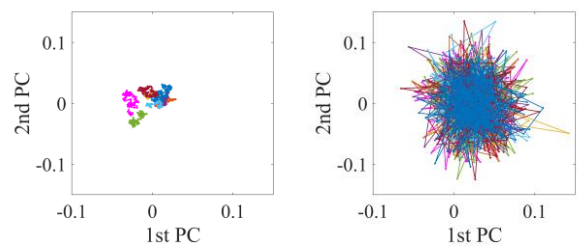

(a) Samples of Model Class 1 (6-para).

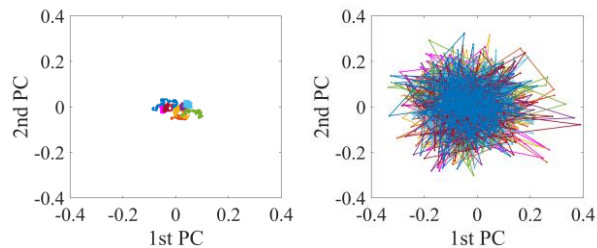

(b) Samples of Model Class 2 (17-para).

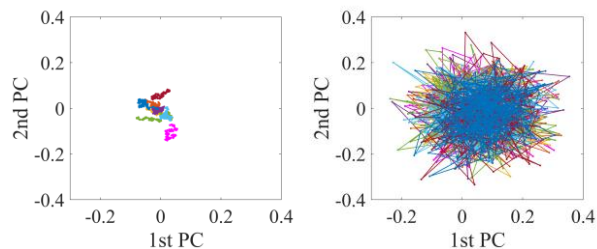

(c) Samples of Model Class 3 (29-para).

**Figure 12.** Sample plots for first two PCs of parameters by HMC (left column) and APM-SGHMC (right column) for three cases (colors mark distinct Markov chains).

As in the previous example, to compare the performance of the different samplers, the negative ELBO (KLD + c), Effective Sample Size (ESS) and sampling efficiency (ESS/h) for both methods on each of the model classes are evaluated and tabulated in Tables 5-7, respectively.

**Table 5.** Sampling performance comparison of Model Class 1 (6-para).

| Methods | KLD + c | ESS | Time (h) | ESS/h |
|---|---|---|---|---|
| HMC | −1009.65 | 6.04 | 38.16 | 0.16 |
| APM-SGHMC | **−1013.57** | **390.8** | 31.76 | **12.30** |

**Table 6.** Sampling performance comparison of Model Class 2 (17-para).

| Methods | KLD + c | ESS | Time (h) | ESS/h |
|---|---|---|---|---|
| HMC | −1033.71 | 4.77 | 75.33 | 0.064 |
| APM-SGHMC | **−1050.87** | **406.6** | 78.31 | **5.19** |

**Table 7.** Sampling performance comparison of Model Class 3 (29-para).

| Methods | KLD + c | ESS | Time (h) | ESS/h |
|---|---|---|---|---|
| HMC | −993.88 | 4.50 | 118.79 | 0.038 |
| APM-SGHMC | **−1034.17** | **107.5** | 130.75 | **0.822** |

As can be seen from the tables, the negative ELBO values (KLD + c) indicate that the distribution of the samples generated by APM-SGHMC fits the posterior PDF better again. As for the sampling efficiency, similar to the previous example, the sampling efficiency of APM-SGHMC is 76, 81 and 21 times that of HMC in 6-para, 17-para and 29-para tasks, respectively. In addition, considering that the ESS calculated by Eq. (54) actually has a lower bound of approximately 1.8, this implies that small ESS values do not decrease proportionally with fewer computational steps. In other words, the sampling efficiency of HMC is actually overestimated. Meanwhile, the relatively strong linear correlation observed in Fig. 11(c) suggests that the 1600-step adaptation phase may not be sufficiently long, resulting in unstable adaptive estimates that adversely affect the ESS of APM-SGHMC in Table 7. Nevertheless, these results show the universal generalization ability of APM-SGHMC among different problems such as with different type of structure.

## 6. CONCLUDING REMARKS

In recent decades, MCMC methods have been extensively studied, with a typical application being the quantification of posterior uncertainties in Bayesian system identification of structural dynamic models. Recently, NN-enhanced MCMC algorithms were developed to improve performance for specific tasks. However, a key challenge is the need for retraining the embedded NNs for new tasks, which is time-consuming and thereby diminishes their competitiveness. As a solution, meta-learning MCMC methods such as AM-SGHMC were developed to reduce training time. Nevertheless, their generalization capability and sampling efficiency are constrained by their simplified NN architectures. For instance, the component-wise NN architectures designed for meta-learning across tasks with varying dimensions lack rotational invariance, making the trained samplers struggle to generalize to the diverse orientations of posterior distributions.

In this paper, a novel meta-learning stochastic simulation approach, called the APM-SGHMC, is developed to enable universal meta-learning capabilities. It achieves rotation-invariance property based on its newly developed adaptive PC transformation technique. Specifically, by reformulating the domain-specific parameters in AM-SGHMC into abstract PCs, APM-SGHMC automatically inherits translation-, scale-, and rotation-invariance from its mathematical structure. This innovation enables generalizable knowledge acquisition of samplers trained on minimalistic tasks, simultaneously overcoming the sampling efficiency limitations inherent in neural network design trade-offs. To accommodate rapid convergence during the burn-in phase of APM-SGHMC sampling, we generalize the adaptivity-precision trade-off in the adaptive PC estimator into an iteration-varying framework and propose a convergence diagnostic and divergence mitigation approach based on affine-invariant statistics, thereby ensuring practical feasibility and robustness.

Two examples involving Bayesian system identification of a multi-story braced-frame building structural model and a bridge structural model subjected to ground motions are used to demonstrate the effectiveness and universal generalization capability of APM-SGHMC. Specifically, in the building example, the adaptive PC meta-learning sampler of APM-SGHMC is first trained on a four-story building model updating task with a typical noise level, and then tested on three generalization tasks involving two-story, four-story and six-story buildings with larger noise levels. The results show that APM-SGHMC achieves both superior sampling fitness and 39, 21 and 19 times, respectively, higher sampling efficiency (Effective Sample Size per hour, ESS/h) than AM-SGHMC, and with 265, 174 and 171 times, respectively, higher efficiency than the classical HMC method. In the bridge example, the APM-SGHMC sampler pre-trained on the building model generalizes to three bridge model identification tasks with 6, 17 and 29 parameters, respectively, without retraining. The results show that APM-SGHMC achieves both significantly superior sampling fitness and 76, 81 and 21 times, respectively, higher sampling efficiency (ESS/h) than HMC. These examples demonstrate that, by reformulating domain-specific parameters into rotation-invariant PCs, APM-SGHMC substantially overcomes the case-by-case limitations of traditional data-driven approaches. It not only attains universal generalization capability, but also achieves substantially boosted sampling fitness and efficiency.

**Declaration of competing interest**

The authors declare that they have no known competing financial interests or personal relationships that could have appeared to influence the work reported in this paper.

**Acknowledgements**

This study was supported by the National Key Research and Development Program of China [Grant No. 2021YFF0501003].

**References**

1 Cheung S. H. and Beck J. L. (2009). Bayesian model updating using Hybrid Monte Carlo Simulation with application to structural dynamics models with many uncertain parameters. Journal of Engineering Mechanics 135: 243–255.
2 Robert, C. P. and Casella, G. (1999). Monte Carlo statistical methods, Springer, New York.
3 Beck, J. L. and Au, S. K. (2002). "Bayesian updating of structural models and reliability using Markov chain Monte Carlo simulation." J. Eng. Mech., 128(2): 380–391.
4 Ching, J., Muto, M. and Beck, J. L. (2006). "Structural model updating and health monitoring with incomplete modal data using Gibbs Sampler." Comput. Aided Civ. Infrastruct. Eng., 21(4): 242–257.
5 Ching, J. and Chen, Y. J. (2007). "Transitional Markov chain Monte Carlo method for Bayesian model updating, model class selection, and model averaging." J. Eng. Mech., 133(7): 816–832.
6 Muto, M. and Beck, J. L. (2008). "Bayesian updating and model class selection for hysteretic structural models using stochastic simulation." J. Vib. Control, 14(1–2): 7–34.
7 Ching, J. and Wang, J. S. (2016). Application of the transitional Markov chain Monte Carlo algorithm to probabilistic site characterization. Engineering Geology; 203:151-167.
8 Wu, S., Angelikopoulos, P., Papadimitriou, C. and Koumoutsakos, P. (2018). Bayesian Annealed Sequential Importance Sampling: An Unbiased Version of Transitional Markov Chain Monte Carlo. ASCE-ASME J Risk Uncertain Eng Syst Part B Mech Eng; 4(1): 011008.

9    Beck J. L. (2010). Bayesian system identification based on probability logic. Struct Control Health Monit; 17(7): 825-847.

10   Huang Y., Shao C., Wu B., Beck J. L. and Li H. (2019). State-of-the-art review on Bayesian inference in structural system identification and damage assessment. Adv Struct Eng; 22(6): 1329-1351.

11   Zhao M., Huang Y., Zhou W. and Li H. (2021). Bayesian uncertainty quantification for guided-wave-based multidamage localization in plate-like structures using Gibbs sampling. Struct Health Monit; 20(6): 3092-3112

12   Yuen K. V., Ching J. and Phoon K. K. (2021). Bayesian Learning Methods for Geotechnical Data. ASCE-ASME J Risk Uncertain Eng Syst, Part A-Civ Eng; 7(1): 02020002.

13   Zhu Z., Au S. K., Li B. and Xie Y. L. (2021). Bayesian operational modal analysis with multiple setups and multiple (possibly close) modes. Mech Syst and Signal Proc; 150: 107261.

14   Jia X., Sedehi O., Papadimitriou C., Katafygiotis L. S. and Moaveni B. (2022). Hierarchical Bayesian modeling framework for model updating and robust predictions in structural dynamics using modal features. Mech Syst and Signal Proc; 170: 108784.

15   Li J., Huang Y. and Asadollahi P. (2021). Sparse Bayesian learning with model reduction for probabilistic structural damage detection with limited measurements. Eng Struct; 247: 113183.

16   Zhou, K., and Tang, J. (2018). Uncertainty quantification in structural dynamic analysis using two-level Gaussian processes and Bayesian inference. J Sound Vibr; 412: 95-115.

17   Wan, H. P., and Ni, Y. Q. (2018). Bayesian modeling approach for forecast of structural stress response using structural health monitoring data. J Struct Eng; 144(9): 04018130.

18   Zhou, K., and Tang, J. (2021). Computational inference of vibratory system with incomplete modal information using parallel, interactive and adaptive Markov chains. J Sound Vibr; 511: 116331.

19   Duane, S., Kennedy, A. D., Pendleton, B. J. and Roweth, D. (1987). Hybrid Monte Carlo. Physics Letters B, 195(2):216–222.

20   Neal, R. M. (2011). MCMC using Hamiltonian dynamics in Handbook of Markov Chain Monte Carlo, 2(11).

21   Ahn, S., Korattikara, A. and Welling, M. (2012). Bayesian posterior sampling via stochastic gradient fisher scoring. arXiv:1206.6380.

22   Patterson S. and Teh, Y. W. (2013). Stochastic gradient Riemannian Langevin dynamics on the probability simplex. In Advances in Neural Information Processing Systems, pp. 3102–3110.

23   Catanach T. A. and Beck J. L. (2017). Bayesian system identification using auxiliary stochastic dynamical systems. International Journal of Nonlinear Mechanics 94: 72–83.

24   Levy, D., Hoffman, M. D. and Sohl-Dickstein, J. (2017). Generalizing Hamiltonian Monte Carlo with Neural Networks. arXiv:1711.09268.

25   Song, J., Zhao, S. and Ermon, S. (2017). A-NICE-MC: Adversarial Training for MCMC. *Advances in neural information processing systems*, 30.

26   Gong, W., Li, Y. and Hernández-Lobato, J. M. (2018). Meta-Learning for Stochastic Gradient MCMC. arXiv:1806.04522.

27   Schmidhuber, J. (1987). Evolutionary principles in self-referential learning, or on learning how to learn: the meta-meta-... hook. PhD thesis, Technische Universität München.

28  Bengio, S., Bengio, Y., Cloutier, J. and Gecsei, J. (1992). On the optimization of a synaptic learning rule. In Conference on Optimality in Biological and Artificial Networks.

29  Naik, D. K. and Mammone R. J. (1992). Meta-neural networks that learn by learning. In International Joint Conference on Neural Networks, v.1, pp. 437–442. IEEE.

30  Thrun, S. and Pratt, L. (1998). Learning to learn. Springer Science & Business Media.

31  Meng, X., Beck, J. L., Huang Y. and Li H. (2025). Adaptive meta-learning stochastic gradient Hamiltonian Monte Carlo simulation for Bayesian updating of structural dynamic models. Comput Meth Appl Mech Eng; 437: 117753.

32  Jaynes, E. T. (2003). Probability Theory: The Logic of Science, Cambridge University Press, 1329–1330.

33  Jaynes, E. T. (1957). Information Theory and Statistical Mechanics, Phys. Rev. 106 (4): 620–630.

34  Kingma, D. P. and Ba, J. (2014). Adam: A Method for Stochastic Optimization. Computer Science.

35  Hotelling, H. (1933). Analysis of a complex of statistical variables into principal components. Journal of educational psychology; 24(6): 417.

36  Jolliffe, I. T. and Jorge C. (2016). Principal component analysis: a review and recent developments. Philosophical transactions of the royal society A: Mathematical, Physical and Engineering Sciences; 374(2065): 20150202.

37  Beck, J. L. and Bernal, D. (2001). A benchmark problem for structural health monitoring. Experimental Techniques 25(3): 49-52.

38  Johnson, E. A., Lam, H. F., Katafygiotis, L. S. and Beck, J. L. (2004). Phase I IASC-ASCE structural health monitoring benchmark problem using simulated data. Journal of Engineering Mechanics, 130(1): 3-15.

39  Yuen, K. V., Au, S. K. and Beck, J. L. (2004). Two-stage structural health monitoring approach for phase I benchmark studies. Journal of Engineering Mechanics, 130(1): 16-33.

40  Ching, J. and Beck, J. L. (2003). Two-step Bayesian structure health monitoring approach for IASC-ASCE phase II simulated and experimental benchmark studies. California Institute of Technology, Earthquake Engineering Research Laboratory.

41  Paszke, A., Gross, S., Massa, F., Lerer, A., Bradbury, J., Chanan, G., ... and Chintala, S. (2019). Pytorch: An imperative style, high-performance deep learning library. Advances in neural information processing systems; 32.

42  Mazzoni, S., McKenna, F., Scott, M.H. and Fenves, G.L. (2006). OpenSees command language manual. Pacific Earthquake Engineering Research (PEER) Center; 264.

43  Ni, P., Han, Q., Du, X. and Cheng, X. (2022). Bayesian model updating of civil structures with likelihood-free inference approach and response reconstruction technique. Mech Syst and Signal Proc; 164: 108204.

44  Zhu, M., McKenna, F. and Scott, M.H. (2018). OpenSeesPy: Python library for the OpenSees finite element framework. SoftwareX; 7: 6-11.

# Appendix A *The essence and a property of the PCD statistical vector*

The statistical vector in Eq. (22) can be rewritten as:

$$\vec{p}(\vec{n}) = \frac{1}{N}\sum_{i=1}^{N}\langle w^{(i)},\vec{n}\rangle w^{(i)} = \frac{1}{N}\sum_{i=1}^{N}w^{(i)}\langle w^{(i)},\vec{n}\rangle$$

$$= \frac{1}{N}\sum_{i=1}^{N}w^{(i)}w^{(i)T}\vec{n} = \left(\frac{1}{N}\sum_{i=1}^{N}w^{(i)}w^{(i)T}\right)\vec{n} \qquad (57)$$

Notably, the covariance formula for the parameter samples $\{w^{(i)}\}_{i=1}^{N}$ is given by:

$$\Sigma = \frac{1}{N}\sum_{i=1}^{N}w^{(i)}w^{(i)T} \qquad (58)$$

Eq. (57) can be further rewritten as:

$$\vec{p}(\vec{n}) = \Sigma\vec{n} \qquad (59)$$

Combining with Eq. (23), i.e., $\vec{n}_{\vec{p}}(\vec{n}) = \frac{\vec{p}(\vec{n})}{|\vec{p}(\vec{n})|}$, the mathematical essence of Eqs. (22)-(23) is a single iteration in the eigen-decomposition of the covariance matrix $\Sigma$ using the power iteration method. Therefore, it can be used to iteratively achieve the PC decomposition of the samples $\{w^{(i)}\}_{i=1}^{N}$.

Since the covariance matrix $\Sigma \in \mathbb{R}^{D\times D}$ is a real symmetric matrix, it has $D$ real eigenvalues $\lambda_1 = \sigma_1^2, \cdots, \lambda_D = \sigma_D^2$, where $\lambda_1$ is the eigenvalue with the largest absolute value (the dominant eigenvalue). Assume that the set of eigenvectors $\{\vec{n}_i\}_{i=1}^{D}$ of $\Sigma$ spans the $\mathbb{R}^D$ space, where $\vec{n}_1$ is an arbitrary unit vector in the 1st PC subspace $S_1$. Then, any unit direction vector $\vec{n}$ in this space can be expressed as:

$$\vec{n} = c_1\vec{n}_1 + \cdots + c_D\vec{n}_D \qquad (60)$$

where $c_1^2 + \cdots + c_D^2 = 1$. Thus:

$$\langle\vec{n},\vec{n}_1\rangle = c_1 \qquad (61)$$

Substituting Eq. (60) into Eq. (59) yields:

$$\vec{p}(\vec{n}) = c_1 \Sigma \vec{n}_1 + \cdots + c_D \Sigma \vec{n}_D = c_1 \lambda_1 \vec{n}_1 + \cdots + c_D \lambda_D \vec{n}_D \tag{62}$$

We then have:

$$\langle \vec{p}(\vec{n}), \vec{n}_1 \rangle = c_1 \lambda_1 \tag{63}$$

$$|\vec{p}(\vec{n})|^2 = c_1^2 \lambda_1^2 + \cdots + c_D^2 \lambda_D^2 \leq (c_1^2 + \cdots + c_D^2) \lambda_1^2 = \lambda_1^2 \tag{64}$$

Combining with Eq. (23), we obtain:

$$\langle \vec{n}_p(\vec{n}), \vec{n}_1 \rangle^2 = \frac{\langle \vec{p}(\vec{n}), \vec{n}_1 \rangle^2}{|\vec{p}(\vec{n})|^2} = \frac{c_1^2 \lambda_1^2}{|\vec{p}(\vec{n})|^2} \geq \frac{c_1^2 \lambda_1^2}{\lambda_1^2} = c_1^2 = \langle \vec{n}, \vec{n}_1 \rangle^2 \tag{65}$$

This corresponds to Eq. (24) in the main text.

Analyzing the inequality condition in Eq. (65), the equality holds if and only if one of the following two conditions is satisfied: (1) $|\vec{p}(\vec{n})|^2 = \lambda_1^2$ or (2) $c_1 = 0$. For condition (1), Eq. (64) reveals that this is equivalent to: $\lambda_i = \lambda_1$ for all $c_i \neq 0$ ($i = 1, \cdots, D$), implying that $\vec{n}$ already lies in $S_1$, i.e., $\vec{n} \in S_1$. For condition (2), the arbitrariness of $\vec{n}_1$ in the subspace $S_1$ implies that $\vec{n}$ is orthogonal to $S_1$. This property in Eq. (65) justifies the use of the statistical vector $\vec{p}(\vec{n})$ in the APM-SGHMC algorithm.

## Appendix B *Bias correction of EMA estimates*

Taking variance estimation as an example, since the initial value $v_0^*$ of the EMA estimate cannot equal the true value $v_t^*$ at each step $t = 1, \cdots, T_{Ada}$, it is necessary to correct the bias introduced by this initial value. The goal is to ensure the expectation of the bias-corrected estimate $\hat{v}_t$ precisely equals the true value $v_t^*$, i.e.:

$$\mathbb{E}[\hat{v}_t] = v_t^* \tag{66}$$

Here, $v_t^*$ is defined as the expectation of the estimate when the initial value equals the true value:

$$v_t^* = \mathbb{E}[v_t | v_0^* = v_t^*] \tag{67}$$

For convenience in subsequent derivation, we first simplify Eq. (38) as:

$$v_t = \beta_{2,t} v_{t-1} + \delta^{(t)} \tag{68}$$

where $\delta^{(t)}$ represents newly introduced statistical terms at step $t$ in Eq. (38) that are independent of $v_0^*$. When $t = 1$, given the initial value $v_0 = v_0^*$, the EMA estimate $v_1$ can be expressed as:

$$v_1 = \beta_{2,1} v_0 + \delta^{(1)} = \left(\prod_{i=1}^{1} \beta_{2,i}\right) v_0^* + c^{(1)} \tag{69}$$

where $c^{(1)}$ is a statistic independent of $v_0^*$. For $t = 2, \cdots, T_{Ada}$, if $v_{t-1}$ satisfies:

$$v_{t-1} = \left(\prod_{i=1}^{t-1} \beta_{2,i}\right) v_0^* + c^{(t-1)} \tag{70}$$

where $c^{(t-1)}$ is independent of $v_0^*$, then substituting into Eq. (68) yields:

$$v_t = \beta_{2,t} \left[\left(\prod_{i=1}^{t-1} \beta_{2,i}\right) v_0^* + c^{(t-1)}\right] + \delta^{(t)}$$

$$= \beta_{2,t} \left(\prod_{i=1}^{t-1} \beta_{2,i}\right) v_0^* + \left(\beta_{2,t} c^{(t-1)} + \delta^{(t)}\right)$$

$$= \left(\prod_{i=1}^{t} \beta_{2,i}\right) v_0^* + c^{(t)} \tag{71}$$

where $c^{(t)} = \beta_{2,t} c^{(t-1)} + \delta^{(t)}$ remains independent of $v_0^*$. Note the consistency between Eqs. (71) and (70), and that when $t = 2$, Eq. (70) reduces to Eq. (69). Thus, Eq. (71) holds for $t = 1, \cdots, T_{Ada}$.

The expectation of $v_t$ is therefore:

$$\mathbb{E}[v_t] = \left(\prod_{i=1}^{t} \beta_{2,i}\right) v_0^* + \mathbb{E}[c^{(t)}] \tag{72}$$

Substituting Eq. (72) into Eq. (67):

$$v_t^* = \left(\prod_{i=1}^{t} \beta_{2,i}\right) v_t^* + \mathbb{E}[c^{(t)}] \tag{73}$$

Eliminating $\mathbb{E}[c^{(t)}]$ by combining Eqs. (72)-(73), we obtain:

$$v_t^* = \frac{\mathbb{E}[v_t] - (\prod_{i=1}^{t} \beta_{2,i}) v_0^*}{1 - \prod_{i=1}^{t} \beta_{2,i}} \tag{74}$$

Consequently, the expectation of the bias-corrected estimate $\hat{v}_t$ in Eq. (40) is:

$$\mathbb{E}[\hat{v}_t] = \mathbb{E}\left[\frac{v_t - (\prod_{i=1}^{t} \beta_{2,i}) v_0^*}{1 - \prod_{i=1}^{t} \beta_{2,i}}\right] = \frac{\mathbb{E}[v_t] - (\prod_{i=1}^{t} \beta_{2,i}) v_0^*}{1 - \prod_{i=1}^{t} \beta_{2,i}} = v_t^* \tag{75}$$

which satisfies the requirement of Eq. (66).

Since the mean estimation process can also be expressed in the form of Eq. (68), the same bias-correction formula applies:

$$\hat{m}_t = \frac{m_t - (\prod_{i=1}^{t} \beta_{1,i}) m_0^*}{1 - \prod_{i=1}^{t} \beta_{1,i}} \tag{76}$$

Noting that $m_0^* = 0$, this simplifies to Eq. (39) in the main text.

## Appendix C *Analysis of the recommended shortest waiting period for the proposed decay-rate sequence*

Take the 1st moment estimate as an example. For the convenience of subsequent derivation, we first simplify Eq. (36) as follows:

$$m_t = \beta_{1,t} m_{t-1} + (1 - \beta_{1,t}) \varepsilon^{(t)} \tag{77}$$

where $\beta_{1,t}$ is defined as $\beta_{1,t} = \frac{\hat{t}_{1,t}-1}{\hat{t}_{1,t}}$ in Eq. (34) and $\varepsilon^{(t)}$ represents the single-batch statistics at step $t$ in Eq. (36). Similar to Eq. (71), the estimate $m_t$ can be organized into the form of a weighted average of the initial value $m_0^*$ and a statistic $m_{stat}^{(t)}$:

$$m_t = \left(\prod_{i=1}^{t} \beta_{1,i}\right) m_0^* + \left(1 - \prod_{i=1}^{t} \beta_{1,i}\right) m_{stat}^{(t)} \tag{78}$$

where $m_{stat}^{(t)}$ is a weighted average of $\{\varepsilon^{(i)}\}_{i=1}^{t}$.

In the proposed three-stage decay-rate sequence controlled by Eq. (43), the role of the waiting period $t_1^b$ is to ensure that the uncertainty of the statistic $m_{stat}^{(t)}$ is low at the end of the first stage, i.e. $t = \hat{t}_{1,min} + t_1^b$, thereby avoiding affecting the second stage, whose role is to steadily reduce the uncertainty to the third stage.

We first focus on the second stage by setting $t_1^b = \hat{t}_{1,min} = 0$. This eliminates the first stage and starts directly from the second stage:

$$\hat{t}_{1,t} = t, \qquad t \leq \hat{t}_{1,max} \tag{79}$$

This setting will make $m_t$ the mean of $\{\varepsilon^{(i)}\}_{i=1}^{t}$ [31]. Considering that $\beta_{1,1} = \frac{\hat{t}_{1,1}-1}{\hat{t}_{1,1}} = 0$, Eq. (78) turns to $m_t = m_{stat}^{(t)}$, such that:

$$m_{stat}^{(t)} = \frac{1}{t}\sum_{i=1}^{t} \varepsilon^{(i)}, \qquad t \leq \hat{t}_{1,max} \tag{80}$$

Assuming that $\{\varepsilon^{(i)}\}_{i=1}^{t}$ are independently and identically distributed (i.i.d.) as a Gaussian distribution $\varepsilon^{(i)} \sim \mathcal{N}(\boldsymbol{\mu}, \boldsymbol{\Sigma})$, the covariance matrix of $m_{stat}^{(t)}$ will be:

$$\boldsymbol{\Sigma}_{stat}^{(t)} = \frac{\boldsymbol{\Sigma}}{t} = \frac{\boldsymbol{\Sigma}}{\hat{t}_{1,t}}, \qquad t \leq \hat{t}_{1,max} \tag{81}$$

In the normal scenario where $\hat{t}_{1,min} > 0$ and $t_1^b > 0$, since $t = \hat{t}_{1,min} + t_1^b$ at the beginning of the second stage, we have $\hat{t}_{1,t} = t - t_1^b = \hat{t}_{1,min}$. Thus, $\boldsymbol{\Sigma}_{stat,target}^{(\hat{t}_{1,min}+t_1^b)} = \frac{\boldsymbol{\Sigma}}{\hat{t}_{1,min}}$ is a suitable reference target.

After analyzing the second stage in detail, we now turn our attention to the first stage under the conditions of $\hat{t}_{1,min} > 0$ and $t_1^b > 0$:

$$\hat{t}_{1,t} = \hat{t}_{1,min}, \qquad t \leq \hat{t}_{1,min} + t_1^b \tag{82}$$

For the convenience of subsequent derivation, we denote the constant decay rate as

$\beta = \frac{\hat{t}_{1,min}-1}{\hat{t}_{1,min}}$. According to the recurrence formula Eq. (77), the following general-term formula can be obtained:

$$\boldsymbol{m}_t = \beta^t \boldsymbol{m}_0^* + (1-\beta)\sum_{i=1}^{t}\beta^{t-i}\boldsymbol{\varepsilon}^{(i)}, \qquad t \le \hat{t}_{1,min} + t_1^b \tag{83}$$

By comparing with Eq. (78), we can obtain:

$$\boldsymbol{m}_{stat}^{(t)} = \frac{(1-\beta)}{(1-\beta^t)}\sum_{i=1}^{t}\beta^{t-i}\boldsymbol{\varepsilon}^{(i)}, \qquad t \le \hat{t}_{1,min} + t_1^b \tag{84}$$

Thus, the covariance matrix of $\boldsymbol{m}_{stat}^{(t)}$ will be:

$$\boldsymbol{\Sigma}_{stat}^{(t)} = \frac{(1-\beta)^2}{(1-\beta^t)^2}\sum_{i=1}^{t}\beta^{2(t-i)}\boldsymbol{\Sigma}$$

$$= \frac{1+\beta^t}{(1-\beta^t)(1+\beta)}\frac{\boldsymbol{\Sigma}}{\hat{t}_{1,min}}, \qquad t \le \hat{t}_{1,min} + t_1^b \tag{85}$$

It can be seen that the ratio coefficient between it and the target value decreases monotonically from $\hat{t}_{1,min}$ with the increase of the time step $t$, converging to $\frac{1}{1+\beta} \approx$ 0.5. Therefore, there exists a $t^*$ such that $\boldsymbol{\Sigma}_{stat}^{(t^*)} = \boldsymbol{\Sigma}_{stat,target}^{(\hat{t}_{1,min}+t_1^b)} = \frac{\boldsymbol{\Sigma}}{\hat{t}_{1,min}}$, which can be obtained by solving:

$$\frac{1+\beta^{t^*}}{(1-\beta^{t^*})(1+\beta)} = 1 \tag{86}$$

After solving, we obtain:

$$t^* = \log_\beta\left(\frac{\beta}{2+\beta}\right) \tag{87}$$

In the left neighborhood of $\beta = 1$, considering $\hat{t}_{1,min} = \frac{1}{1-\beta}$, Eq. (87) can be approximated by:

$$t^* \approx \log(3)\,\hat{t}_{1,min} + \frac{2}{3} - \frac{\log(3)}{2}$$

$$\approx 1.0987\hat{t}_{1,min} + 0.1174$$

$$\approx 1.1\hat{t}_{1,min} \tag{88}$$

Since $\hat{t}_{1,min}$ and $t^*$ are usually integers, a tolerance of 1 is adopted. Thus, Eq. (88) is approximately an upper bound of Eq. (87). With this as a reference, we suggest that the waiting period $t_1^b$ satisfies:

$$t_1^b \geq t^* - \hat{t}_{1,min} \approx 0.1\hat{t}_{1,min} \tag{89}$$

and the same applies to $t_2^b$ and $t_3^b$.

## Appendix D *NN architecture and training setup*

The architecture of the meta-strategy NN is shown in Fig. 13, which can be partitioned into three parts according to the dashed-line boxes.

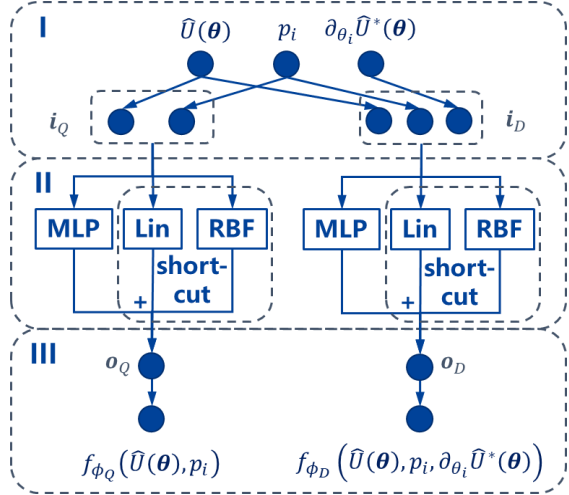

**Figure 13.** Schematic of the meta-strategy NN architecture.

Parts I and III serve as additional input/output processing modules to adapt the NN for training. In part I, the inputs $\widehat{U}(\boldsymbol{\theta})$, $p_i$ and $\partial_{\theta_i}\widehat{U}^*(\boldsymbol{\theta})$ are transformed into NN inputs $\boldsymbol{i}_U$, $\boldsymbol{i}_{p_i}$ and $\boldsymbol{i}_{G_i}$ by functions:

$$i_U = \log\left[\left(ReLU(\widehat{U}(\boldsymbol{\theta}) + 1)\right)^2 + e - 1\right] - 1 \tag{90}$$

$$\boldsymbol{i}_{p_i} = 3\,Sigmoid\left(\frac{p_i}{10}\right) - 1.5 \tag{91}$$

$$\boldsymbol{i}_{G_i} = 3\,Sigmoid\left(\frac{\partial_{\theta_i}\widehat{U}^*(\boldsymbol{\theta})}{30}\right) - 1.5 \tag{92}$$

Here, $ReLU(\cdot)$ and $Sigmoid(\cdot)$ are two widely-used activation functions in NNs. In Part III, the NN outputs $\boldsymbol{o}_Q$ and $\boldsymbol{o}_D$ are converted into function outputs by:

$$f_{\phi_Q}(\widehat{U}(\boldsymbol{\theta}), p_i) = M_Q\,Sigmoid(5\boldsymbol{o}_Q) \tag{93}$$

$$f_{\phi_D}\left(\widehat{U}(\boldsymbol{\theta}), p_i, \partial_{\theta_i}\widehat{U}^*(\boldsymbol{\theta})\right) = M_D\,Sigmoid(5\boldsymbol{o}_D) \tag{94}$$

where $M_Q = 100$ and $M_D = 30$, given the step-size $\eta = \sqrt{0.001}$, as suggested in [31].

Regarding Part II, each embedded NN adopts a parallel architecture that integrates an MLP, a Lin (linear-transform) shortcut, and an RBF shortcut, as shown in the following equation:

$$\boldsymbol{o} = \text{NN}(\boldsymbol{i}) = \text{MLP}(\boldsymbol{i}) + \text{Lin}(\boldsymbol{i}) + \text{RBF}(\boldsymbol{i}) \tag{95}$$

Each MLP has 3 hidden layers, with 10 units in each hidden layer. The Leaky ReLU function is employed as the activation function for each unit in the hidden layers. Each RBF shortcut consists of 10 Gaussian basis function units.

For the training process, $K_0 = 64$ parallel chains are simulated for 50 epochs, with each epoch comprising 10 sub-epochs. After each epoch, states $\boldsymbol{z} = (\boldsymbol{\theta}, \boldsymbol{p})$ of the Markov chains are re-initialized using replay techniques with a replay probability of 0.6, that is, 60% of the states are re-initialized using the states simulated earlier.

For each sub-epoch, the sampler is simulated 90 steps and then updated using the Adam optimizer with a learning rate of 0.002 and exponential decay rates $(\beta_1, \beta_2)_{\text{Adam}} = (0.9, 0.999)$. Every $T = 15$ steps, $K = 10$ chains are randomly selected to accumulate the gradient for backpropagation. The RBF shortcuts are only accumulated in the first 25 epochs and the Lin shortcuts are only accumulated in the last 9 sub-epochs of the first 25 epochs.

The adaptive estimates are updated only during the last 5 sub-epochs of the first epoch and the last 9 sub-epochs of the first 2-40 epochs. The parameters for the updates are $(t_1^b, \hat{t}_{1,min}, \hat{t}_{1,max})_\theta = (100,100,1000)$ , $(t_2^b, \hat{t}_{2,min}, \hat{t}_{2,max})_\theta = (5000,1000,2000)$ , $(t_3^b, \hat{t}_{3,min}, \hat{t}_{3,max})_\theta = (200,1000,2000)$ and $(\beta_1, \beta_2)_U = (0.99, 0.998)$ for parameter samples $\theta$ and potential energy $U(\theta)$.